\documentclass[11pt]{amsart}
\usepackage{amsmath}
\usepackage{mathtools}
\usepackage{amssymb}
\usepackage{enumerate}
\usepackage{slashed}
\usepackage{graphicx}
\usepackage{newlfont}
\usepackage{amsrefs}
\usepackage{comment}
\usepackage[normalem]{ulem}
\usepackage{mathtools}
\usepackage{accents}
\usepackage{leftidx}
\usepackage{cancel}
\usepackage[marginpar=18mm,top=28mm,bottom=28mm,right=22mm,left=22mm]{geometry}

\numberwithin{equation}{section}

\theoremstyle{plain}
\newtheorem{theorem}{Theorem}
\newtheorem{lemma}[theorem]{Lemma}
\newtheorem{prop}[theorem]{Proposition}

\theoremstyle{definition}

\theoremstyle{remark}

\newcommand{\R}{\mathbb{R}}

\begin{document}
\title[Balancing Static Vacuum Black Holes]{Balancing Static Vacuum Black Holes with Signed Masses in 4 and 5 Dimensions}

\author[Khuri]{Marcus Khuri}
\address{Department of Mathematics\\
Stony Brook University\\
Stony Brook, NY 11794, USA}
\email{khuri@math.sunysb.edu}

\author[Weinstein]{Gilbert Weinstein}
\address{Physics Department and Department of Mathematics\\
Ariel University\\
Ariel, 40700, Israel}
\email{gilbertw@ariel.ac.il}

\author[Yamada]{Sumio Yamada}
\address{Department of Mathematics\\
Gakushuin University\\
Tokyo 171-8588, Japan}
\email{yamada@math.gakushuin.ac.jp}

\thanks{M. Khuri acknowledges the support of NSF Grant DMS-1708798, and Simons Foundation Fellowship 681443. S. Yamada acknowledges the support of JSPS KAKENHI Grant JP17H01091.}

\begin{abstract}
We construct a new set of asymptotically flat, static vacuum solutions to the Einstein equations in dimensions 4 and 5, which may be interpreted as a superposition of positive and negative mass black holes. The resulting spacetimes are axisymmetric in 4-dimensions and bi-axisymmetric in 5-dimensions, and are regular away from the negative mass singularities, for instance conical singularities are absent along the axes. In 5-dimensions, the topologies of signed mass black holes used in the construction may be either spheres $S^3$ or rings $S^1 \times S^2$; in particular, the negative mass static black ring solution is introduced. A primary observation that facilitates the superposition is the fact that, in Weyl-Papapetrou coordinates, negative mass singularities arise as overlapping singular support for a particular type of Green's function. Furthermore, a careful analysis of conical singularities along axes is performed, and formulas are obtained for their propagation across horizons, negative mass singularities, and corners. The methods are robust, and may be used to construct a multitude of further examples. Lastly, we show that balancing does not occur between any two signed mass black holes of the type studied here in 4 dimensions, while in 5 dimensions two-body balancing is possible.
\end{abstract}
\maketitle

\section{Introduction}
\label{sec1} \setcounter{equation}{0}
\setcounter{section}{1}

According to the classical black hole uniqueness (no hair) theorem \cite{BuntingMasood1,Israel1,KhuriWoolgar,Robinson1}, a finite number of vacuum black holes cannot be held in asymptotically flat static equilibrium if there is more than one. In fact, the only asymptotically flat static vacuum black hole solution is the Schwarzschild spacetime. It should be noted that the typical assumption of asymptotic flatness can be relaxed \cite{Anderson1,Reiris1}, and that the result also holds in higher dimensions \cite{GibbonsIdaShiromizu1}. On the other hand, an infinite number of black holes may be configured so that the total force experienced by each constituent vanishes, yielding a regular static vacuum solution. These solutions are constructed in 4 and 5 dimensions \cite{korotkinnicolai,KhuriWeinsteinYamadaJHEP,myers1987}, and are asymptotically Kasner. A further non-asymptotically flat example of multiple vacuum black holes held in static equilibrium is given in \cite{AV}, where the mechanism responsible for regularity of the solution is an external gravitational field. The balancing of multiple static black holes may also be achieved with charge. Indeed, if the black holes carry an appropriate amount of charge, then electromagnetic repulsion and gravitational attraction balance to produce the static Majumdar-Papapetrou solutions \cite{Majumdar,Papapetrou}. These are the only regular multi-black hole solutions of the static electro-vacuum equations \cite{ChruscielTod} which are asymptotically flat, and they admit generalizations to higher dimensions \cite{myers1987}.

In this paper we pursue an alternative method to balance multiple static vacuum black holes, namely by employing negative mass naked singularities. In particular, we construct a new set of regular, asymptotically flat, static vacuum spacetimes in 4 and 5 dimensions that arise as a superposition of positive and negative mass Schwarzschild and black ring solutions. It is also shown that
in order for such balancing to occur in 4D, the superposition must consist of at least three signed mass black holes. Surprisingly, however, balancing is possible in 5D with only two signed mass black holes.
The mechanism by which these solutions are produced is robust, and we indicate how it may be utilized to obtain a variety of further examples. Those that are given here can exhibit positive, negative, or zero ADM (total) mass.

Negative mass in general relativity was first discussed in some detail by Bondi \cite{Bondi} (see also \cite{BonnorSwaminarayan}), where the basic mechanics were described and a two-body problem was studied. Assuming that negative masses obey the equivalence principle, the gravitational interaction between signed masses results in motion that is consistent with intuition from Newton's laws: two positive masses attract, two negative masses repel, and one positive and one negative mass accelerate in the same direction with the negative mass following the positive mass. As described by Bonnor \cite{Bonnor}, this last property can result in `runaway motion' with constant acceleration, if the two masses are of the same magnitude. In contrast, similarly to electrostatics, if a negative mass is situated on a line at the midpoint between two positive masses of the same value, and the magnitude of the negative mass is a fourth of each positive mass, then the system is in gravitational equilibrium. It is this simple observation that motivates the balancing of static vacuum black holes exhibited in this work. Although we do not attempt to find applications of these new solutions here, it should be noted that negative masses have been investigated within the context of wormholes \cite{Bronnikov,EllisH,MorrisThorneYurtsever}, dark energy/matter \cite{Farnes,Socas-Navarro}, spacecraft propulsion \cite{Forward,Landis}, and the Penrose inequality \cite{BrayJauregui}.

All of the static vacuum solutions discussed in this paper will be axisymmetric in 4-dimensions and  bi-axisymmetric in 5-dimensions, so that the isometry group of the associated spacetimes contains the subgroup $\mathbb{R}\times U(1)$ and $\mathbb{R}\times U(1)^2$, respectively. The spacetime metric on the domains of outer communication may be expressed in Weyl-Papapetrou coordinates \cite{Harmark}, in which the metric coefficients are determined by a single harmonic function in 4D and two harmonic functions in 5D. More generally, in the 5-dimensional setting, a harmonic map from $\mathbb{R}^3 \rightarrow SL(3,\mathbb{R})/SO(3)$ may be used to obtain solutions having lens space $L(p,q)$ horizon topology, however here we will restrict attention to the spheres $S^3$ and rings $S^1 \times S^2$. In \cite{BachWeyl}, Bach and Weyl analyzed 4D axisymmetric static vacuum black holes and showed that when more than one horizon is present a conical singularity must appear on the axis between horizons. This conical singularity, or strut along the axis, may then be interpreted as a force holding the black holes in equilibrium. The Bach-Weyl solutions are constructed by superpositioning Green's functions on $\mathbb{R}^3$ with singular support along intervals of the $z$-axis. The singular support intervals are referred to as \textit{axis rods}, whereas the remaining intervals on the $z$-axis are associated with individual horizons and are called \textit{horizon rods}.

In Section \ref{sec2} below, we make the observation that the singularity present in the 4D negative mass Schwarzschild solution, when described in Weyl-Papapetrou coordinates, corresponds to the overlapping of singular support intervals for two Green's functions. It is this fact that allows us to build solutions to the static vacuum equations which consist of any finite combination of Schwarzschild black holes, and Schwarzschild \textit{negative mass singularities (NMS)}. These solutions, when expressed in Weyl-Papapetrou coordinates, exhibit three different types of intervals along the $z$-axis: horizon rods, NMS rods, and axis rods. In this context there are only two possible types of singularities, namely negative mass singularities at NMS rods and conical singularities along axis rods. Conical singularities may be characterized as having an angle deficit or angle surplus when the logarithmic angle defect
$b=\log\left(2\pi\cdot\mathrm{radius}/\mathrm{circumference}\right)$ is positive or negative, respectively. This may be interpreted as a force
exerted by the axis rod on its adjacent horizons and/or NMSs, and in 4D the relation \cite{Weinstein1990} is given by
\begin{equation}\label{1}
F=\frac{1}{4}\left(e^{-b}-1\right).
\end{equation}
When the force is zero on all axis rods, the solution will be referred to as \textit{balanced}.

\begin{theorem}\label{thmfirst}
Any finite collection of signed mass Schwarzschild black holes may be superpositioned to form a 4-dimensional asymptotically flat, axisymmetric, static vacuum spacetime. These solutions are regular away from the negative mass singularities, except perhaps for conical singularities on the axes.
\begin{itemize}
\item [(i)] (Two Bodies) Any of these configurations consisting of two signed mass black holes must possess a conical singularity. In particular, if $m_1$ and $m_2$ are the signed masses then the force associated with the axis rod separating them satisfies: $F>0$ if $\mathrm{sign}(m_1) = \mathrm{sign}( m_2)$, and $F<0$ if $\mathrm{sign}(m_1) \neq \mathrm{sign} (m_2)$.

\item [(ii)] (Three Bodies) There exist balanced configurations consisting of one negative mass and two positive mass black holes, as well as one positive mass and two negative mass black holes. The former solutions have positive ADM mass, while the latter solutions have negative ADM mass.
\end{itemize}
\end{theorem}

This result is established by computing the propagation of angle defects across horizons and NMSs, resulting in explicit formulas for the force in terms of the two masses and separation distance between the horizon/NMS rods. The conical singularity formulas are of independent interest, and are presented in Section \ref{sec3}. These formulas with minor modifications will be applicable also in the rotating case. Previous work addressing the issue of conical singularities in this 4D situation includes  \cite{LiTian,Weinstein1994}. In addition, an example of a self-accelerating Bondi dipole, without conical singularities and made of a positive and negative mass Schwarzschild black hole, was given in \cite{EmparanGarriga}.
Part (i) of the theorem may be considered as a generalization of the analysis presented by Bach and Weyl \cite{BachWeyl}, who considered only the case of positive masses. To the best of the authors' knowledge, the balancing of positive and negative mass black holes as described in part (ii) seems to be a new, although perhaps not unexpected, phenomena. It is anticipated that a plethora of further balanced examples may be obtained with the same methods for more than three black holes. We also point out that the sign of the force in this theorem is consistent with the intuition obtained from the mechanics of signed mass particles, as well as the interpretation that $F$ measures the force exerted by an axis rod on its neighboring black holes. For instance, consider the case of a solution having a single positive mass and single negative mass black hole. Without the force arising from the central axis rod, the two masses would accelerate in the direction of the positive mass due to gravity. According to part (i) the axis rod force satisfies $F<0$. When acting on the positive mass, the negative force results in an attraction of the mass towards the rod, and similarly a repulsion of the negative mass away from the rod. Thus, intuitively, the accelerations resulting from the gravitational and axis rod force negate each other so that the two signed mass black holes remain in equilibrium, albeit in a conically singular spacetime.

In 5-dimensions, bi-axisymmetric static black hole horizon cross-sections may be of three topological types \cite{HollandsYazadjiev,KhuriWeinsteinYamada}, namely a sphere $S^3$, a ring $S^1 \times S^2$, or a lens space $L(p,q)$ where $p$ and $q$ are coprime integers. Emparan and Reall \cite{EmparanReall} have analyzed the vacuum solutions, including those with multiple horizons, and have found that except for the Schwarzschild-Tangherlini black hole, those that are asymptotically flat all possess conical singularities. Indeed, this is consistent with the higher dimensional version of static black hole uniqueness \cite{GibbonsIdaShiromizu1}. The negative mass Schwarzschild-Tangherlini solution behaves analogously to its 4D counterpart, in that it has a point singularity (within a constant time slice) which when expressed in Weyl-Papapetrou coordinates becomes an interval on the $z$-axis arising from the overlap of the singular support of two Green's functions; this will be referred to as a \textit{spherical NMS}. Furthermore, a new type of 5D NMS will be discussed here that arises from the black ring family of solutions, namely a static ring of negative mass. As with the spherical NMS, this is obtained by the overlapping of singular support sets associated with certain Green's functions. The singularity is again topologically a point, which corresponds to an interval on the $z$-axis in Weyl-Papapetrou coordinates; this will be referred to as a \textit{ring NMS}. Moreover, a neighborhood of the ring NMS in a constant time slice is foliated by rings $S^1 \times S^2$ that collapse onto the singularity. Using the characterization in terms of Green's functions, we may superposition any number of spherical horizons/NMSs and ring horizons/NMSs to obtain asymptotically flat static vacuum solutions consisting of signed mass black holes having $S^3$ and $S^1\times S^2$ topologies. As in 4D, conical singularities on the axis rods may be interpreted as a force similar to \eqref{1}. However, as we will show, in 5D this force is not necessarily constant along each axis rod, see Section \ref{force} for details. A careful analysis of conical singularity propagation across horizons, NMSs, and corners (the intersection point of two axis rods) leads to formulas for these type of singularities in terms of masses and rod lengths.
Here too as in 4D, this analysis should generalize to the rotating case.
With these formulas we produce examples of balanced solutions consisting of three signed mass black holes having spherical and ring elements. Unexpectedly, in contrast to 4-dimensions, we also find balanced two-body solutions consisting of one negative mass and one positive mass black hole involving both spheres and rings.

\begin{theorem}\label{thmsecond}
Any finite collection of 5-dimensional signed mass Schwarzschild-Tangherlini and signed mass static black ring solutions, may be superpositioned to form an bi-axisymmetric, static vacuum spacetime. These solutions are regular away from the negative mass singularities, except perhaps for conical singularities on the axes.
\begin{itemize}
\item [(i)] (Two Bodies) There exist asymptotically flat balanced configurations consisting of one spherical horizon and one spherical negative mass singularity, as well as one spherical horizon and one ring negative mass singularity (negative mass black Saturn). The former solutions have zero ADM mass, while the latter solutions can have ADM mass of any sign.

\item [(ii)] (Three Bodies) There exist asymptotically flat balanced configurations consisting of one spherical negative mass singularity and two spherical horizons, as well as one spherical and one ring negative mass singularity together with one ring horizon. The former solutions have positive ADM mass, while the latter solutions have negative ADM mass.
\end{itemize}
\end{theorem}

This paper is organized as follows. In Section \ref{sec2} we discuss the necessary background material in the 4D setting, and make the observation that negative mass Schwarzschild solutions arise from the overlapping of singular support for Green's functions. Section \ref{sec3} is dedicated to the analysis of conical singularities and their propagation across horizons and NMSs in 4D, and the proof of Theorem \ref{thmfirst} is given in Section \ref{sec4}. For the 5D setting, background material and a discussion of negative mass static black holes appears in Section \ref{sec5}, while the analysis of conical singularities is carried out in Section \ref{sec6}. Finally, Theorem \ref{thmsecond} is proved in Section \ref{sec7}.

\subsection*{Acknowledgements}
The authors would like to thank Roberto Emparan for helpful comments.

\section{4D Background and the Negative Mass Schwarzschild Spacetime}
\label{sec2} \setcounter{equation}{0}
\setcounter{section}{2}

Consider the domain of outer communication $\mathcal{M}^4$
of a static axisymmetric 4-dimensional spacetime. Under reasonable hypotheses \cite{ChruscielCostaHeusler}, the orbit space $\mathcal{M}^4/[\mathbb{R}\times U(1)]$ is homeomorphic to the right half plane $\{(\rho,z)\mid \rho>0\}$. The spacetime metric may then be expressed in Weyl-Papapetrou coordinates
\begin{equation}\label{metricg}
g = - e^{-u}\rho^2 dt^2 + e^u  d \phi^2 +  e^{2 \alpha} (d \rho^2 + dz^2),
\end{equation}
where $\partial_{\phi}$ is the generator of the $U(1)$ symmetry with $\phi\in[0,2\pi)$. The vacuum Einstein equations in this setting \cite{Weinstein1990} reduce to
the Laplace equation for $u$, and a set of quadrature equations for $\alpha$, on $\mathbb{R}^3\setminus \{z-\text{axis}\}$ parameterized by the cylindrical coordinates $(\rho,z,\varphi)$, namely
\begin{equation}\label{einsteineq}
\Delta u=0,\quad\quad\quad \alpha_{\rho}=\frac{\rho}{4}\left[u_{\rho}^2-u_{z}^2 -\frac{2}{\rho} u_{\rho}\right],\quad\quad\quad
\alpha_{z}=\frac{\rho}{2}\left[u_{\rho} u_{z}-\frac{1}{\rho}u_{z}\right].
\end{equation}
Notice that the integrability conditions for the $\alpha$ equations correspond to the harmonicity of $u$. Furthermore, the $z$-axis is decomposed into a sequence of intervals called \textit{rods}, that are denoted by $\{\Gamma_l\}_{l\in I}$ for some index set $I$. There are two types of rods, those on which $|\partial_{\phi}|$ vanishes are referred to as \textit{axis rods}, and those on which $|\partial_t|$ vanishes are referred to as \textit{horizon rods}. Later we will introduce a third type of rod based on negative mass singularities. The intersection point of an axis rod with a horizon rod is called a \textit{pole}.

According to the above description, any solution of the axisymmetric static vacuum Einstein equations may be obtained by specifying a harmonic function on $\mathbb{R}^3$ that has appropriate blow-up behavior on portions of the $z$-axis.
Unlike $u$, the behavior of the function $\alpha$ cannot be prescribed along axes, and thus conical singularities may form on axis rods when constructing the spacetime metric $g$ in \eqref{metricg}. A conical singularity at point $(0,z_0)$ on an axis rod $\Gamma_{l}$, may be determined from the \textit{angle defect} $\theta\in(-\infty,2\pi)$ associated with the 2-dimensional cone formed by the orbits of $\partial_{\phi}$ over the line $z=z_0$. More precisely
\begin{equation}\label{csingularity}
\frac{2\pi}{2\pi-\theta}=\lim_{\rho\rightarrow 0}\frac{2\pi\cdot\mathrm{Radius}}{\mathrm{Circumference}}
=\lim_{\rho\rightarrow 0}
\frac{\int_{0}^{\rho}\sqrt{e^{2\alpha}}}{e^{u/2}}
=\lim_{\rho\rightarrow 0} \rho e^{\alpha-u/2}.
\end{equation}
It follows from \eqref{einsteineq} that this quantity is constant along each axis rod. Moreover, the absence of a conical singularity corresponds to a zero logarithmic angle defect. In this situation, the metric is smoothly extendable across the axis, as may be checked with a change to Cartesian coordinates. We will denote the \textit{logarithmic angle defect} $\log\left(\frac{2\pi}{2\pi-\theta}\right)$, on the axis rod $\Gamma_l$, by $b_l$.  The conical singularity on $\Gamma_l$ is then said to exhibit an \textit{angle deficit} if $b_l>0$, and an \textit{angle surplus} if $b_l<0$. The sign of the logarithmic angle defect determines the character of the force associated with the axis rod. Namely, as computed in \cite[pg. 921]{Weinstein1990}, this force is constant at each point of the rod and is given by
\begin{equation}
F=\frac14 \left(e^{-b_l} -1\right).
\end{equation}

\subsection{Green's functions on $\R^3$}

The function $u$ used to construct static vacuum spacetimes with horizons and negative mass singularities will be sums of Green's functions having singular support on intervals of the $z$-axis. For instance, on an axis rod $\Gamma_l$ we have $|\partial_{\phi}|=0$ so that $u(x)\rightarrow-\infty$ as $x\rightarrow \Gamma_l$. Blow-up behavior, although with a different rate, will also characterize negative mass singularity rods, while $u$ remains bounded on horizon rods. Here we detail the type of Green's functions that arise, and introduce notation.

Consider an interval $\Gamma$ on the $z$-axis. The Green's function associated to $\Gamma$, with blow-up modeled by $2\log\rho$, will be denoted by $G_{\Gamma}$. This function satisfies the distributional equation
\begin{equation}
\Delta G_{\Gamma}=2\delta_{\Gamma},
\end{equation}
where $\delta_{\Gamma}$ is the Dirac delta distribution for $\Gamma$. Thus, $G_{\Gamma}$ may be viewed as the potential for a constant positive charge distribution along $\Gamma$. The simplest example has singular support on the entire $z$-axis, namely $G_{\{\ \!\!\!z\mathrm{-axis}\}}=2\log \rho$.
Moreover, the finite interval Green's functions may be built from the following two semi-infinite interval Green's functions:
\begin{equation}
G_{\{z<a\}}=\log\left[\sqrt{\rho^2 +(z-a)^2} +(z-a)\right],\quad\quad G_{\{b<z\}}=\log\left[\sqrt{\rho^2 +(z-b)^2} -(z-b)\right].
\end{equation}
In particular, we have two ways to express a finite interval Green's function
\begin{equation}
G_{\{a<z<b\}}= G_{\{z<b\}}-G_{\{z<a\}}=\log\left[\frac{\sqrt{\rho^2 +(z-b)^2} +(z-b)}
{\sqrt{\rho^2 +(z-a)^2} +(z-a)}\right],
\end{equation}
or equivalently
\begin{equation}
G_{\{a<z<b\}}= G_{\{z>a\}}-G_{\{z>b\}}=\log\left[\frac{\sqrt{\rho^2 +(z-a)^2} -(z-a)}
{\sqrt{\rho^2 +(z-b)^2} -(z-b)}\right].
\end{equation}
Notice that
\begin{equation}\label{ggg}
G_{\{a<z<b\}}=2\log\rho +O(1)\quad\text{ as }\quad (\rho,z)\rightarrow (0,z),\quad a<z<b,
\end{equation}
and this function is bounded on the complement of the interval.

\subsection{The signed mass Schwarzschild solution}

In order to motivate the Weyl-Papapetrou coordinate presentation of the negative mass Schwarzschild solution, we first recall the corresponding presentation for the positive mass solution. In this case, the domain of outer communication has topology $\mathbb{R}\times \left(\mathbb{R}^3 \setminus \text{Ball}\right)$, and in Schwarzschild coordinates the spacetime metric takes the form
\begin{equation}\label{scmetric}
g_{S}= - \left( 1 - \frac{2m}{\bar{r}} \right) dt^2 + \left( 1 - \frac{2m}{\bar{r}} \right)^{-1} d\bar{r}^2 + \bar{r}^2 (d \theta^2 + \sin^2 \theta d \phi^2),
\end{equation}
where the mass $m>0$ and $\bar{r}\geq 2m$, $0\leq\theta\leq \pi$, $0\leq\phi<2\pi$.
In general, for stationary axisymmetric spacetimes, the cylindrical radius $\rho$ of Weyl-Papapetrou coordinates is obtained from the determinant of the Killing $t\phi$-portion of the metric. The $z$ coordinate is then chosen as a harmonic conjugate, so that $(\rho,z)$ form isothermal coordinates on the orbit space $\mathcal{M}^4 /[\mathbb{R}\times U(1)]$. It follows that
\begin{equation}
\rho = \bar{r}\sin \theta  \sqrt{1 - \tfrac{2m}{\bar{r}}} ,\quad\quad\quad
z  = (\bar{r}-m) \cos \theta .
\end{equation}
The spacetime metric $g_S$ then takes the form \eqref{metricg}, where
\begin{equation}
u=2 \log \rho - \log \left( 1- \frac{2m}{\bar{r}} \right)
=2\log\rho - G_{\{-m < z < m\}}=G_{\{z>m\}}+G_{\{z<-m\}}
\end{equation}
and
\begin{equation}
\alpha=\frac{1}{2}\log\left[\frac{(r_+ + r_-)^2 -4m^2}{4 r_+ r_-}\right]+\frac{1}{2}u-\log\rho,
\end{equation}
with $r_{\pm}=\sqrt{\rho^2 +(z\mp m)^2}$ denoting the Euclidean distance to the poles. From this we find that the $z$-axis is broken into two axis rods $\Gamma_{-}=(-\infty,-m)$ and $\Gamma_{+}=(m,\infty)$ where $|\partial_{\phi}|$ vanishes, and one horizon rod $\Gamma_{h}=(-m,m)$ where $|\partial_t|$ vanishes which corresponds to the surface $\bar{r}=2m$ in Schwarzschild coordinates.

Consider now the negative mass Schwarzschild solution. The topology of the domain of outer communication is the same as the positive mass case, although geometrically it may be preferable to think of a point being removed from $\mathbb{R}^3$, instead of a ball, since the singularity has zero area. The metric expression \eqref{scmetric} still remains valid, although here the mass parameter $m<0$ and the areal radius extends to the origin $\bar{r}>0$. Notice that there is no longer a coordinate singularity at $\bar{r}=2m$, and there is no event horizon so that $\bar{r}=0$ is a naked singularity. The formulas for the change to Weyl-Papapetrou coordinates have the same form
\begin{equation}
\rho = \bar{r}\sin \theta  \sqrt{1 + \tfrac{2|m|}{\bar{r}}} ,\quad\quad\quad
z  = (\bar{r}+|m|) \cos \theta ,
\end{equation}
as does the harmonic function metric coefficient
\begin{equation}
u=G_{\{z>-|m|\}}+G_{\{z<|m|\}},
\end{equation}
and the function $\alpha$. There is, however, a change in the rod structure which will require the introduction of a new type of rod.

\begin{figure}
\includegraphics[width=.9\textwidth]{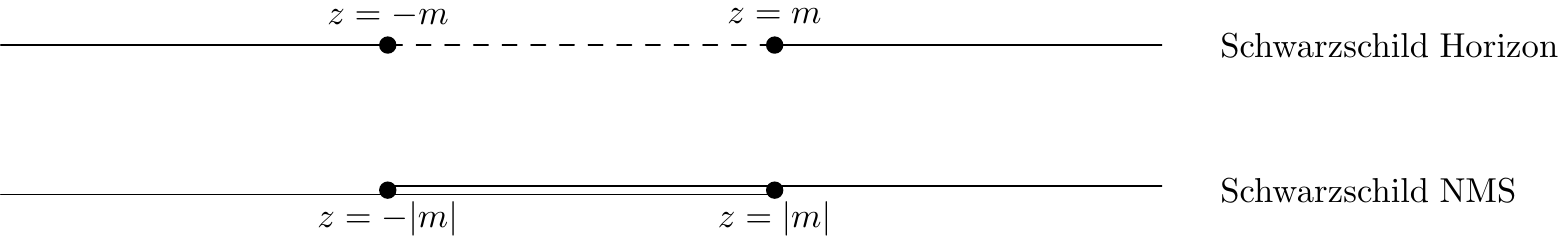}
\caption{Rod structures for Schwarzschild metrics of positive and negative mass. Here, as in future rod diagrams, the $z$-axis is drawn horizontally and the dotted lines indicate horizon rods, while the overlapping portions indicate NMS rods.}
\label{4d-schw-zas1}
\end{figure}

The rod structure is determined by the asymptotic behavior of $u$ upon approach to the $z$-axis. Observe that the singular support sets for the Green's functions that comprise $u$, overlap on the interval $\Gamma_s =(-|m|,|m|)$. As above, let $\Gamma_{-}=(-\infty,-|m|)$ and $\Gamma_{+}=(|m|,\infty)$, then it follows from \eqref{ggg} and \eqref{einsteineq} that as $\rho\rightarrow 0$ we have
\begin{equation}\label{2.16}
u=
\begin{cases}
2\log\rho +O(1) & \text{if $z\in \Gamma_{\pm}$} \\
4\log\rho +O(1) & \text{if $z\in\Gamma_s$}
\end{cases},
\quad\quad\quad
\alpha=
\begin{cases}
O(1) & \text{if $z\in \Gamma_{\pm}$} \\
2\log\rho +O(1) & \text{if $z\in\Gamma_s$}
\end{cases}.
\end{equation}
This shows that the rods $\Gamma_{\pm}$ behave as typical axis rods, whereas $\Gamma_s$ does not exhibit the characteristic of either an axis or horizon rod.
We will therefore refer to $\Gamma_s$, and more generally any rod of an axisymmetric static spacetime near which $u=4\log\rho +O(1)$, as a \textit{negative mass singularity (NMS) rod}. These intervals of the $z$-axis correspond to single points in time slices of the spacetime, and may be viewed as the overlap of two neighboring axis rods, see Figure \ref{4d-schw-zas1}.

\section{Conical Singularities in Dimension 4}
\label{sec3} \setcounter{equation}{0}
\setcounter{section}{3}

In this section we analyze the propagation of conical singularities across horizons and NMSs, for axisymmetric static vacuum solutions. Formulas for the difference of the logarithmic angle defect between two neighboring axis rods will be given in terms of renormalized values for $u$ at the poles. Let $z_2 < z_s <z_n < z_1$ denote values on the $z$-axis, and consider a sequence of two axis rods $\Gamma_1 =(z_n , z_1)$ to the north, and $\Gamma_2 =(z_2, z_s)$ to the south, that border a horizon rod $\Gamma_h =(z_s, z_n)$ with poles $N$, $S$ located at $z_n$, $z_s$. According to \eqref{csingularity}, and the fact that angle defects are constant along axis rods, the relation between the two logarithmic angle defects may be computed at any two points along these two axes. In particular, if $\varepsilon>0$ is a small parameter, we will use points $N_{\varepsilon}=(0,z_n +\varepsilon)$ and $S_{-\varepsilon}=(0,z_s -\varepsilon)$ to find
\begin{align}\label{mainform}
\begin{split}
b_1 -b_2 =& \lim_{\rho\rightarrow 0}\left(\alpha-\frac{1}{2}u+\log\rho\right)\Big|_{z=z_n +\varepsilon} -\lim_{\rho\rightarrow 0}\left(\alpha-\frac{1}{2}u+\log\rho\right)\Big|_{z=z_s - \varepsilon}\\
=& \frac{1}{2}\left(\bar{u}(S_{-\varepsilon})-\bar{u}(N_{\varepsilon})\right)
+\alpha(N_{\varepsilon})-\alpha(S_{-\varepsilon}),
\end{split}
\end{align}
where $\bar{u}=u-2\log\rho$.

In order to evaluate the right-hand side of \eqref{mainform}, expansions for $u$ and $\alpha$ at the poles are needed to isolate the singular parts. From the Green's function analysis of the previous section, we find that
\begin{equation}\label{renormalizationh}
u=\log\left( r_n -(z-z_n)\right)+\hat{u}_n\quad \text{ near } N,\quad\quad
u=\log\left( r_s +(z-z_s)\right)+\hat{u}_s\quad \text{ near } S,
\end{equation}
for some smooth functions $\hat{u}_n$ and $\hat{u}_s$, where $r_{n/s}=\sqrt{\rho^2 +(z-z_{n/s})^2}$ is the Euclidean distance to $N$ and $S$. Furthermore using these expressions and \cite[(5.15)]{KhuriWeinsteinYamadaJHEP}, the expansions for $\alpha$ are found to be
\begin{equation}
\alpha=-\frac{1}{2}\log r_n +\frac{(z-z_n)}{2r_n}\left(\hat{u}_n -\hat{u}_n(N)\right)+\hat{c}_n +O(\rho^2)\quad\text{ near }N,
\end{equation}
\begin{equation}
\alpha=-\frac{1}{2}\log r_s -\frac{(z-z_s)}{2r_s}\left(\hat{u}_s -\hat{u}_s(S)\right)+\hat{c}_s +O(\rho^2)\quad\text{ near }S,
\end{equation}
where $\hat{c}_n$ and $\hat{c}_s$ are constants. Next note that the renormalizations $\bar{u}$ for axis rods, and $\hat{u}_{n/s}$ for poles are related. More precisely
\begin{align}\label{3.5}
\begin{split}
\bar{u}(N_{\varepsilon})=&\lim_{\rho\rightarrow 0}\left(u-2\log\rho\right)|_{z=z_n +\varepsilon}\\
=& \hat{u}_n (N_{\varepsilon})+\lim_{\rho\rightarrow 0}\log\left[\frac{\sqrt{\rho^2 +(z-z_n)^2}-(z-z_n)}{\rho^2}\right]_{z=z_n +\varepsilon}\\
=&\hat{u}_n (N_{\varepsilon})+\log\left[\frac{1}{2(z-z_n)}\right]_{z=z_n +\varepsilon}\\
=&\hat{u}_n (N_{\varepsilon}) -\log(2\varepsilon),
\end{split}
\end{align}
and similarly
\begin{equation}
\bar{u}(S_{-\varepsilon})=\hat{u}_s(S_{-\varepsilon})-\log(2\varepsilon).
\end{equation}
It follows that the first term on the right-hand side of \eqref{mainform} becomes
\begin{equation}\label{p098}
 \frac{1}{2}\left(\bar{u}(S_{-\varepsilon})-\bar{u}(N_{\varepsilon})\right)
 =\frac{1}{2}\left(\hat{u}_s(S)-\hat{u}_n(N)\right)+O(\varepsilon).
\end{equation}

In order to treat the remaining terms of \eqref{mainform}, let $N_{-\varepsilon}=(0,z_n -\varepsilon)$ and $S_{\varepsilon}=(0,z_s +\varepsilon)$ be reflections across the poles of $N_{\varepsilon}$ and $S_{-\varepsilon}$. These reflected points lie inside the horizon rod, where we may better use the quadrature equation \eqref{einsteineq} to obtain relations between $\alpha$ and $u$. Notice that the expansions for $\alpha$ imply
\begin{equation}
\alpha(N_{\varepsilon})-\alpha(N_{-\varepsilon})
=\frac{1}{2}\left(\hat{u}_n(N_{\varepsilon})-\hat{u}_n(N)\right)
+\frac{1}{2}\left(\hat{u}_n(N_{-\varepsilon})-\hat{u}_n(N)\right)=O(\varepsilon),
\end{equation}
and similarly
\begin{equation}
\alpha(S_{\varepsilon})-\alpha(S_{-\varepsilon})=O(\varepsilon).
\end{equation}
Therefore
\begin{align}\label{-=}
\begin{split}
\alpha(N_{\varepsilon})-\alpha(S_{-\varepsilon})=& \left[\alpha(N_{\varepsilon})-\alpha(N_{-\varepsilon})\right]
+\left[\alpha(N_{-\varepsilon})-\alpha(S_{\varepsilon})\right]
+\left[\alpha(S_{\varepsilon})-\alpha(S_{-\varepsilon})\right]\\
=&\alpha(N_{-\varepsilon})-\alpha(S_{\varepsilon})+O(\varepsilon).
\end{split}
\end{align}
Next observe that since $u$ is regular on the horizon rod, \eqref{einsteineq} along the horizon yields
\begin{equation}\label{-}
\alpha(N_{-\varepsilon})-\alpha(S_{\varepsilon})=\int_{z_s +\varepsilon}^{z_n -\varepsilon}\alpha_{z}(0,z)dz=-\frac{1}{2}\int_{z_s +\varepsilon}^{z_n -\varepsilon}u_{z}(0,z)dz=\frac{1}{2}\left(u(S_{\varepsilon})-u(N_{-\varepsilon})\right).
\end{equation}
Moreover, the expansions for $u$ near the poles give
\begin{equation}\label{+}
u(S_{\varepsilon})-u(N_{-\varepsilon})
=\hat{u}_s(S_{\varepsilon})-\hat{u}_n(N_{-\varepsilon}).
\end{equation}
Putting \eqref{p098}, \eqref{-=}, \eqref{-}, and \eqref{+} altogether, and letting $\varepsilon\rightarrow 0$, produces the desired propagation formula for conical singularities across horizons
\begin{equation}\label{horizonc}
b_1 -b_2 =\hat{u}_s(S)-\hat{u}_n(N).
\end{equation}

We will now obtain a formula analogous to \eqref{horizonc} for NMSs. Consider, as above, a sequence of three rods consisting of a single NMS rod $\Gamma_s$ bordered to the north by $\Gamma_1$ and to the south by $\Gamma_2$. The location of the rods along the $z$-axis, and the notation for relevant quantities will be the same as above, except that the horizon rod $\Gamma_h$ is replaced by the NMS rod $\Gamma_s$. As in the previous argument, evaluation of the right-hand side of \eqref{mainform} requires appropriate expansions for $u$ and $\alpha$ at the poles. More precisely
\begin{equation}\label{renormalizations}
u=2\log\rho +\log\left( r_n +(z-z_n)\right)+\tilde{u}_n\quad \text{near } N,\quad\text{ }\text{ }
u=2\log\rho+\log\left( r_s -(z-z_s)\right)+\tilde{u}_s\quad \text{near } S,
\end{equation}
where $\tilde{u}_n$ and $\tilde{u}_s$ are smooth functions. Furthermore, if we write $u=2\log\rho+w$ then \eqref{einsteineq} shows that
\begin{equation}
\alpha_{\rho}
=w_{\rho}+\frac{\rho}{4}\left(w_{\rho}^2 -w_z^2-\frac{2}{\rho}w_{\rho}\right),\quad\quad\quad
\alpha_{z}
=w_{z}+\frac{\rho}{2}\left(w_{\rho}w_z -\frac{1}{\rho}w_{z}\right).
\end{equation}
The computation \cite[(5.15)]{KhuriWeinsteinYamadaJHEP} can then be used to find
\begin{equation}
\alpha=\log\left(r_n+(z-z_n)\right)+\tilde{u}_n-\frac{1}{2}\log r_n -\frac{(z-z_n)}{2r_n}\left(\tilde{u}_n -\tilde{u}_n(N)\right)+\tilde{c}_n +O(\rho^2)\quad\text{ near }N,
\end{equation}
\begin{equation}
\alpha=\log\left(r_s -(z-z_s)\right)+\tilde{u}_s-\frac{1}{2}\log r_s +\frac{(z-z_s)}{2r_s}\left(\tilde{u}_s -\tilde{u}_s(S)\right)+\tilde{c}_s +O(\rho^2)\quad\text{ near }S,
\end{equation}
for some constants $\tilde{c}_n$ and $\tilde{c}_s$. Next observe that the relation between renormalizations is given by
\begin{equation}
\bar{u}(N_{\varepsilon})=\lim_{\rho\rightarrow 0}\left(u-2\log\rho\right)|_{z=z_n +\varepsilon}
=\lim_{\rho\rightarrow 0}\left[\log\left(r_n +(z-z_n)\right)+\tilde{u}_n\right]_{z=z_n +\varepsilon}=\log(2\varepsilon)+\tilde{u}_n(N_{\varepsilon}),
\end{equation}
and similarly
\begin{equation}
\bar{u}(S_{-\varepsilon})=\log(2\varepsilon)+\tilde{u}_s(S_{-\varepsilon}).
\end{equation}
Thus, the first term on the right-hand side of \eqref{mainform} becomes
\begin{equation}\label{p0981}
 \frac{1}{2}\left(\bar{u}(S_{-\varepsilon})-\bar{u}(N_{\varepsilon})\right)
 =\frac{1}{2}\left(\tilde{u}_s(S)-\tilde{u}_n(N)\right)+O(\varepsilon).
\end{equation}

The function $\alpha$ is not finite along the NMS rod, as it was along the horizon rod.
Therefore, the method of reflecting the points $N_{\varepsilon}$ and $S_{-\varepsilon}$ across the poles to aid with the computation of $\alpha$ terms in \eqref{mainform}, must be modified. Towards this end, define $\tilde{N}_{-\varepsilon}=(\varepsilon_0,z_n -\varepsilon)$ as well as $\tilde{S}_{\varepsilon}=(\varepsilon_0,z_s +\varepsilon)$, where $\varepsilon_0 >0$ is an additional small parameter. The expansions for $\alpha$ then imply
\begin{align}
\begin{split}
\alpha(N_{\varepsilon})-\alpha(\tilde{N}_{-\varepsilon})
=&\log(2\varepsilon)+\tilde{u}_n(N_{\varepsilon})-\frac{1}{2}\log\varepsilon
-\frac{1}{2}\left(\tilde{u}_n(N_{\varepsilon})-\tilde{u}_n(N)\right)
-\log\left(\sqrt{\varepsilon_0^2 +\varepsilon^2}-\varepsilon\right)\\
&-\tilde{u}_n(\tilde{N}_{-\varepsilon})
+\frac{1}{2}\log\sqrt{\varepsilon_0^2 +\varepsilon^2}-\frac{1}{2\sqrt{2}}\left(\tilde{u}_n(\tilde{N}_{-\varepsilon})-
\tilde{u}_n(N)\right),
\end{split}
\end{align}
and similarly
\begin{align}
\begin{split}
\alpha(\tilde{S}_{\varepsilon})-\alpha(S_{-\varepsilon})
=&\log\left(\sqrt{\varepsilon_0^2 +\varepsilon^2}-\varepsilon\right)+\tilde{u}_s(\tilde{S}_{\varepsilon})
-\frac{1}{2}\log\sqrt{\varepsilon_0^2 +\varepsilon^2}+\frac{1}{2\sqrt{2}}\left(\tilde{u}_s(\tilde{S}_{\varepsilon})-
\tilde{u}_s(S)\right)\\
&-\log(2\varepsilon)-\tilde{u}_s(\tilde{S}_{-\varepsilon})+\frac{1}{2}\log\varepsilon
+\frac{1}{2}\left(\tilde{u}_s(S_{-\varepsilon})-\tilde{u}_s(S)\right),
\end{split}
\end{align}
hence
\begin{equation}
\left[\alpha(N_{\varepsilon})-\alpha(\tilde{N}_{-\varepsilon})\right]
+\left[\alpha(\tilde{S}_{\varepsilon})-\alpha(S_{-\varepsilon})\right]
=O(\varepsilon+\varepsilon_0).
\end{equation}
It follows that
\begin{align}\label{-=1}
\begin{split}
\alpha(N_{\varepsilon})-\alpha(S_{-\varepsilon})=& \left[\alpha(N_{\varepsilon})-\alpha(\tilde{N}_{-\varepsilon})\right]
+\left[\alpha(\tilde{N}_{-\varepsilon})-\alpha(\tilde{S}_{\varepsilon})\right]
+\left[\alpha(\tilde{S}_{\varepsilon})-\alpha(S_{-\varepsilon})\right]\\
=&\alpha(\tilde{N}_{-\varepsilon})-\alpha(\tilde{S}_{\varepsilon})
+O(\varepsilon+\varepsilon_0).
\end{split}
\end{align}
This difference in $\alpha$ may be transformed into a difference in $u$ with the help of \eqref{einsteineq} and \eqref{2.16}. Namely, if $z\in \Gamma_s$ is held fixed then as $\rho\rightarrow 0$ we have
\begin{equation}
\alpha_z =\frac{\rho}{2}\left[u_{\rho}u_z -\frac{1}{\rho}u_z\right]
=\frac{3}{2}u_z+O(\rho).
\end{equation}
Therefore if $\varepsilon>0$ is held fixed as $\varepsilon_0 \rightarrow 0$, then
\begin{align}\label{-1}
\begin{split}
\alpha(\tilde{N}_{-\varepsilon})-\alpha(\tilde{S}_{\varepsilon})=&\int_{z_s +\varepsilon}^{z_n -\varepsilon}\alpha_{z}(\varepsilon_0,z)dz\\
=&\frac{3}{2}\int_{z_s +\varepsilon}^{z_n -\varepsilon}u_{z}(\varepsilon_0,z)dz +O(\varepsilon_0)\\
=&\frac{3}{2}\left(u(\tilde{N}_{-\varepsilon})-u(\tilde{S}_{\varepsilon})\right)
+O(\varepsilon_0)\\
=&\frac{3}{2}\left(\tilde{u}_n(N)-\tilde{u}_s(S)\right)
+O(\varepsilon+\varepsilon_0).
\end{split}
\end{align}
Combining \eqref{p0981}, \eqref{-=1}, \eqref{-1}, and letting $\varepsilon_0\rightarrow 0$ before $\varepsilon\rightarrow 0$, yields the desired propagation formula for conical singularities across an NMS
\begin{equation}\label{horizonss}
b_1 -b_2 =\tilde{u}_n(N)-\tilde{u}_s(S).
\end{equation}
We now record what has been shown.

\begin{prop}\label{conicalprop}
Consider the domain of outer communication of a 4-dimensional, asymptotically flat, axisymmetric, static vacuum spacetime, and let $u$ be the corresponding potential of Weyl-Papapetrou coordinates \eqref{metricg}. Let $\Gamma_1$ and $\Gamma_2$ be two axis rods, which  surround either a horizon rod or a NMS rod with north and south poles denoted by $N$ and $S$. If $b_1$ and $b_2$ are the associated logarithmic angle defects of these axis rods, and $\hat{u}$, $\tilde{u}$ are the renormalizations of $u$ given by
\eqref{renormalizationh}, \eqref{renormalizations} respectively, then
\begin{equation}\label{horizonconicform}
b_1 -b_2 =\hat{u}_s(S)-\hat{u}_n(N)\quad \text{ across a horizon}, \text{ }\quad\quad
b_1 -b_2 =\tilde{u}_n(N)-\tilde{u}_s(S)\quad\text{ across a NMS}.
\end{equation}
\end{prop}

\section{Balancing of Signed Masses in 4 Dimensions and the Proof of Theorem \ref{thmfirst}}
\label{sec4} \setcounter{equation}{0}
\setcounter{section}{4}

In this section we will establish Theorem \ref{thmfirst}. The first statement of this result concerns existence for superpositions of positive and negative mass black hole solutions. Indeed, according to the description in terms of Green's functions presented in Section \ref{sec2}, we may define the potential function $u$ as a linear combination of such Green's functions for any configuration of signed mass Schwarzschild black holes strung along the $z$-axis. With $u$, one may solve the quadrature equations \eqref{einsteineq} for $\alpha$. The resulting spacetime, with metric described in Weyl-Papapetrou coordinates \eqref{metricg}, is then
an asymptotically flat and axisymmetric solution of the static vacuum Einstein equations. These solutions are regular away from the negative mass singularities, except perhaps for conical singularities on the axes.

\subsection{Two signed masses are unbalanced}

Here we address Theorem \ref{thmfirst} (i), and show that no configuration consisting of the superposition of two signed mass Schwarzschild solutions is balanced. In particular, an explicit formula for the force associated with the axis between the masses is given and shown to be nonzero in all cases.

We begin by constructing the general rod structure for two signed masses. Consider a sequence of three axis rods $\Gamma_i$, $i=1,2,3$, with $\Gamma_1$ lying to the north (positive $z$ direction) of $\Gamma_2$, and $\Gamma_2$ lying to the north of $\Gamma_3$, where $\Gamma_1$ and $\Gamma_3$ are semi-infinite and $\Gamma_2$ is finite. Between $\Gamma_1$ and $\Gamma_2$ is a signed mass rod $\Gamma_{h/s}^1$ (a horizon or NMS), and between $\Gamma_2$ and $\Gamma_3$ is another signed mass rod $\Gamma_{h/s}^2$. The north and south poles of the signed mass rods will be labeled $N_j =(0,n_j)$, $S_j =(0,s_j)$, $j=1,2$ in the $\rho z$-plane, with $s_2 < n_2 < s_1 < n_1$. Thus, the total rod structure consists of 5 rods and 4 pole points. The associated potential function for this rod structure is given by
\begin{equation}
u=2\log\rho -\mathrm{sign}(m_1) G_{\Gamma_{h/s}^1}-\mathrm{sign}(m_2)G_{\Gamma_{h/s}^2},
\end{equation}
where $m_1$ and $m_2$ are the (Komar) masses of each horizon/NMS, $\mathrm{sign}(m_j)=\pm 1$ depending on whether $m_j$ is positive or negative, and the Green's functions are
\begin{equation}
G_{\Gamma_{h/s}^1}=\log\left[\frac{r_{S_1}-(z-s_1)}{r_{N_1}-(z-n_1)}\right],\quad\quad\quad
G_{\Gamma_{h/s}^2}=\log\left[\frac{r_{N_2}+(z-n_2)}{r_{S_2}+(z-s_2)}\right],
\end{equation}
with $r_{N_j}$ and $r_{S_j}$ representing the Euclidean distance to $N_j$ and $S_j$, respectively.

We will treat first the case in which $\Gamma_{h/s}^1$ is a horizon rod, that is $\mathrm{sign}(m_1)>0$. In order to calculate the propagation of conical singularities across this horizon rod via Proposition \ref{conicalprop}, it is necessary to compute the renormalization for $u$ near the poles. Observe that
\begin{align}
\begin{split}
\hat{u}_{n_1}=& u-\log\left(r_{N_1}-(z-n_1)\right)\\
=&2\log\rho-\log\left[\frac{r_{S_1}-(z-s_1)}{r_{N_1}-(z-n_1)}\right]
-\mathrm{sign}(m_2)G_{\Gamma_{h/s}^2}-\log\left(r_{N_1}-(z-n_1)\right)\\
=&2\log\rho -\log\left(r_{S_1}-(z-s_1)\right)-\mathrm{sign}(m_2)G_{\Gamma_{h/s}^2}\\
=&\log\left(2(z-s_1)\right)-\mathrm{sign}(m_2)G_{\Gamma_{h/s}^2} +O(\rho^2)
\end{split}
\end{align}
if $z>s_1$, so that
\begin{equation}
\hat{u}_{n_1}(N_1)=\log\left(2(n_1-s_1)\right)
-\mathrm{sign}(m_2)G_{\Gamma_{h/s}^2}(N_1).
\end{equation}
Similarly
\begin{align}
\begin{split}
\hat{u}_{s_1}=& u-\log\left(r_{S_1}+(z-s_1)\right)\\
=&2\log\rho-\log\left[\frac{r_{S_1}-(z-s_1)}{r_{N_1}-(z-n_1)}\right]
-\mathrm{sign}(m_2)G_{\Gamma_{h/s}^2}-\log\left(r_{S_1}+(z-s_1)\right)\\
=&2\log\rho +\log\left(r_{N_1}-(z-n_1)\right)
-\log\left(r_{S_1}^2-(z-s_1)^2\right)-\mathrm{sign}(m_2)G_{\Gamma_{h/s}^2}\\
=&\log\left(r_{N_1}-(z-n_1)\right)-\mathrm{sign}(m_2)G_{\Gamma_{h/s}^2},
\end{split}
\end{align}
so that
\begin{equation}
\hat{u}_{s_1}(S_1)=\log\left(2(n_1-s_1)\right)
-\mathrm{sign}(m_2)G_{\Gamma_{h/s}^2}(S_1).
\end{equation}
Hence, Proposition \ref{conicalprop} implies that the difference of conical singularities on the axes $\Gamma_1$ and $\Gamma_2$ is given by
\begin{equation}\label{c1}
b_1 -b_2 =-\mathrm{sign}(m_2)\left(G_{\Gamma_{h/s}^2}(S_1)
-G_{\Gamma_{h/s}^2}(N_1)\right)=-\mathrm{sign}(m_2)
\log\left[\frac{(s_1 -n_2)(n_1 -s_2)}{(s_1 -s_2)(n_1 -n_2)}\right].
\end{equation}

Consider now the case in which $\Gamma_{h/s}^2$ is a NMS rod, that is $\mathrm{sign}(m_1)<0$. Imitating the above computation for the appropriate renormalization, we find that
\begin{align}
\begin{split}
\tilde{u}_{n_1}=& u-\log\left(r_{N_1}+(z-n_1)\right)-2\log\rho\\
=&-2\log\rho +\log\left(r_{S_1}-(z-s_1)\right) -\mathrm{sign}(m_2)G_{\Gamma_{h/s}^2}\\
=&-\log\left(2(z-s_1)\right)-\mathrm{sign}(m_2)G_{\Gamma_{h/s}^2} +O(\rho^2)
\end{split}
\end{align}
if $z>s_1$, so that
\begin{equation}
\tilde{u}_{n_1}(N_1)=-\log\left(2(n_1-s_1)\right)
-\mathrm{sign}(m_2)G_{\Gamma_{h/s}^2}(N_1).
\end{equation}
Similarly
\begin{align}
\begin{split}
\tilde{u}_{s_1}=& u-\log\left(r_{S_1}-(z-s_1)\right)-2\log\rho\\
=&-\log\left(r_{N_1}-(z-n_1)\right)-\mathrm{sign}(m_2)G_{\Gamma_{h/s}^2},
\end{split}
\end{align}
so that
\begin{equation}
\tilde{u}_{s_1}(S_1)=-\log\left(2(n_1-s_1)\right)
-\mathrm{sign}(m_2)G_{\Gamma_{h/s}^2}(S_1).
\end{equation}
Proposition \ref{conicalprop} then implies implies that
\begin{equation}\label{c2}
b_1 -b_2 =-\mathrm{sign}(m_2)\left(G_{\Gamma_{h/s}^2}(N_1)
-G_{\Gamma_{h/s}^2}(S_1)\right)=-\mathrm{sign}(m_2)
\log\left[\frac{(n_1 -n_2)(s_1 -s_2)}{(n_1 -s_2)(s_1 -n_2)}\right].
\end{equation}

The two semi-infinite axes $\Gamma_1$ and $\Gamma_3$ may be assumed to be free of conical singularities, that is $b_1=b_3=0$. This is due to the fact that $\alpha$ is defined only up to addition of a constant, and this constant may be chosen to relieve any conical singularity present on $\Gamma_1$. Furthermore, by integrating the appropriate flux along an arbitrarily large curve in the asymptotic end of the $\rho z$-plane, it can be shown that the angle defects of the semi-infinite rods agree, see \cite[Section 6]{KhuriWeinsteinYamadaP} for details. The expressions \eqref{c1} and \eqref{c2} then give the angle defect for $\Gamma_2$, the axis rod separating the two masses. It is instructive to evaluate this angle defect in terms of masses, and the length of separation between them. To do this note, as is shown in Section \ref{massec1}, that if $m_j$ is the Komar mass of the signed mass rod $\Gamma_{h/s}^j$ then this rod has length $2|m_j|$. Moreover, we will denote the length of the separating axis rod $\Gamma_2$ by $\ell$. The pole points may then be taken to be
\begin{equation}
s_2 =-|m_2|,\quad\quad n_2 =|m_2|, \quad\quad s_1 =\ell+|m_2|,\quad\quad
n_1 =\ell +|m_2|+2|m_1|.
\end{equation}
Therefore, in the case that $\Gamma_{h/s}^1$ is a horizon \eqref{c1} yields
\begin{equation}
b_2 =\mathrm{sign}(m_2)\log\left[\frac{\ell(\ell +2m_1 +2|m_2|)}{(\ell+2m_1)(\ell+2|m_2|)}\right],
\end{equation}
and when $\Gamma_{h/s}^1$ is a NMS \eqref{c2} produces
\begin{equation}
b_2 =\mathrm{sign}(m_2)\log\left[\frac{(\ell+2|m_1|)(\ell+2|m_2|)}
{\ell(\ell +2|m_1| +2|m_2|)}\right].
\end{equation}
The force may now be found according to \eqref{1}.

\begin{prop}\label{force4d}
The superposition of two 4-dimensional signed mass Schwarzschild black holes cannot be held in static equilibrium without a conical singularity. In particular, any static vacuum configuration obtained by superpositioning two Schwarzschild solutions of masses $m_1, m_2 \in \mathbb{R}\setminus\{0\}$, must exhibit a nonzero force along the axis separating the two masses. If $\ell$ denotes the (Euclidean) length of the separating axis rod, then this force is given by
\begin{equation}\label{forceformula4}
F=
\begin{cases}
\frac{|m_1 m_2|}{\ell (\ell+2|m_1|+2|m_2|)}  &   \text{ if } m_1 m_2 >0\\
-\frac{|m_1 m_2|}{(\ell +2|m_1|)(\ell +2|m_2|)} & \text{ if } m_1 m_2 <0
\end{cases}.
\end{equation}
\end{prop}

This shows that two masses of the same sign, yields an axis rod with a positive force that repels them. Note that although the force is repellant, the direction of the resulting acceleration depends on the sign of the masses. Moreover, masses of opposite sign produce an axis rod with a negative, or attractive, force. These forces are, however, balanced by the gravitational force so that the system remains in equilibrium. The computation of the force in the case of two positive masses was known to Bach and Weyl \cite{BachWeyl}, and the inability to balance two signed mass black holes was discussed by Bondi \cite{Bondi}.

The formula \eqref{forceformula4} shows that when the two masses are separated by arbitrarily large distances ($\ell\rightarrow\infty$), the force asymptotes to the inverse square law of Newtonian gravity, as should be expected. On the other hand, when $\ell\rightarrow 0$, the force between two masses of opposite sign does not blow-up. In fact, it limits to the value $F=-1/4$, regardless of the magnitude of the two masses. This new observation, of which there is no Newtonian analogue, seems to have gone unnoticed in the study of negative mass, and deserves further investigation.

\subsection{Balancing three signed masses}

\begin{figure}
	\includegraphics[width=.9\textwidth]{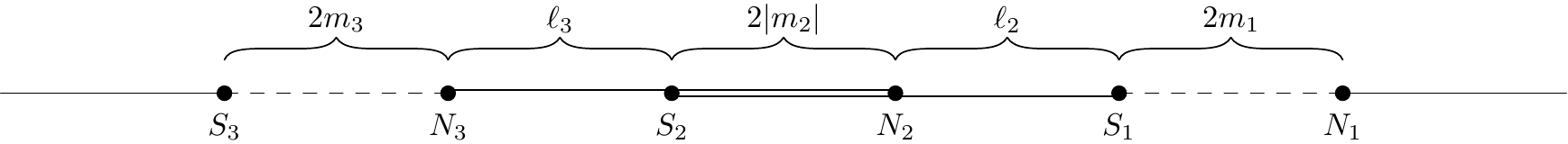}
	\caption{Rod structure for the superposition of three signed masses.}
	\label{4d-SZS}
\end{figure}

Here we will establish Theorem \ref{thmfirst} (ii). Consider a sequence of four axis rods $\Gamma_i$, $i=1,2,3,4$, with the semi-infinite rods $\Gamma_1$, $\Gamma_4$ lying to the north (positive $z$ direction) and south of the others, respectively, and with $\Gamma_2$ lying to the north of $\Gamma_3$.  Between $\Gamma_1$ and $\Gamma_2$ is a horizon rod $\Gamma_{h}^1$, between $\Gamma_2$ and $\Gamma_3$ is a NMS rod $\Gamma_{s}^2$, and between $\Gamma_3$ and $\Gamma_4$ is a horizon rod $\Gamma_h^3$. The north and south poles of the horizon/NMS rods will be labeled $N_j =(0,n_j)$, $S_j =(0,s_j)$, $j=1,2,3$ in the $\rho z$-plane, with $s_3<n_3<s_2 < n_2 < s_1 < n_1$. Thus, the total rod structure consists of 7 rods and 6 pole points, see Figure \ref{4d-SZS}. The associated potential function for this rod structure is given by
\begin{equation}
u=2\log\rho -G_{\Gamma_{h}^1}+G_{\Gamma_{s}^2}-G_{\Gamma_h^3},
\end{equation}
where the Green's functions are
\begin{equation}
G_{\Gamma_{h}^1}=\log\left[\frac{r_{S_1}-(z-s_1)}{r_{N_1}-(z-n_1)}\right],\quad
G_{\Gamma_{s}^2}=\log\left[\frac{r_{N_2}+(z-n_2)}{r_{S_2}+(z-s_2)}\right],\quad
G_{\Gamma_{h}^3}=\log\left[\frac{r_{N_3}+(z-n_3)}{r_{S_3}+(z-s_3)}\right].
\end{equation}

We will calculate the propagation of conical singularities across this horizon rod $\Gamma_h^1$ via Proposition \ref{conicalprop}. This requires computing the renormalization for $u$ near the poles. Namely
\begin{align}
\begin{split}
\hat{u}_{n_1}=& u-\log\left(r_{N_1}-(z-n_1)\right)\\
=&2\log\rho-\log\left[\frac{r_{S_1}-(z-s_1)}{r_{N_1}-(z-n_1)}\right]
+G_{\Gamma_{s}^2}-G_{\Gamma_h^3}-\log\left(r_{N_1}-(z-n_1)\right)\\
=&2\log\rho -\log\left(r_{S_1}-(z-s_1)\right)+G_{\Gamma_{s}^2}-G_{\Gamma_h^3}\\
=&\log\left(2|z-s_1|\right)+G_{\Gamma_{s}^2}-G_{\Gamma_h^3} +O(\rho^2)
\end{split}
\end{align}
if $z>s_1$, so that
\begin{equation}
\hat{u}_{n_1}(N_1)=\log\left(2(n_1-s_1)\right)
+G_{\Gamma_{s}^2}(N_1)-G_{\Gamma_h^3}(N_1).
\end{equation}
Similarly
\begin{align}
\begin{split}
\hat{u}_{s_1}=& u-\log\left(r_{S_1}+(z-s_1)\right)\\
=&2\log\rho-\log\left[\frac{r_{S_1}-(z-s_1)}{r_{N_1}-(z-n_1)}\right]
+G_{\Gamma_{s}^2}-G_{\Gamma_h^3}-\log\left(r_{S_1}+(z-s_1)\right)\\
=&2\log\rho +\log\left(r_{N_1}-(z-n_1)\right)
-\log\left(r_{S_1}^2-(z-s_1)^2\right)+G_{\Gamma_{s}^2}-G_{\Gamma_h^3}\\
=&\log\left(r_{N_1}-(z-n_1)\right)+G_{\Gamma_{s}^2}-G_{\Gamma_h^3},
\end{split}
\end{align}
so that
\begin{equation}
\hat{u}_{s_1}(S_1)=\log\left(2(n_1-s_1)\right)
+G_{\Gamma_{s}^2}(S_1)-G_{\Gamma_h^3}(S_1).
\end{equation}
Hence, Proposition \ref{conicalprop} implies that the difference of conical singularities on the axes $\Gamma_1$ and $\Gamma_2$ is given by
\begin{align}\label{conicalbalance}
\begin{split}
b_1 -b_2 =&\left[G_{\Gamma_s^2}(S_1)-G_{\Gamma_s^2}(N_1)\right]
-\left[G_{\Gamma_{h}^3}(S_1)-G_{\Gamma_h^3}(N_1)\right]\\
=&\log\left[\frac{(s_1 -n_2)(n_1 -s_2)(s_1 -s_3)(n_1 -n_3)}{(s_1 -s_2)(n_1 -n_2)
(s_1 -n_3)(n_1 -s_3)}\right].
\end{split}
\end{align}

As in the previous subsection, the two semi-infinite axes $\Gamma_1$ and $\Gamma_4$ may be assumed to be free of conical singularities, that is $b_1=b_4=0$. In order to evaluate the angle defect $b_2$ in terms of masses, and the length of separation between them let $m_1,m_3>0$, $m_2 <0$ denote the Komar masses of the signed mass rods $\Gamma_{h}^1$, $\Gamma_h^3$, $\Gamma_s^2$ so that these rods have lengths $2m_1$, $2m_3$, $2|m_2|$. Moreover let $\ell_2$, $\ell_3$ be the lengths of the axis rods $\Gamma_2$, $\Gamma_3$ separating the masses. The pole points may then be labelled
\begin{equation}
s_3 =-m_3, \quad\quad n_3 =m_3, \quad\quad s_2=\ell_3 +m_3,\quad\quad n_2 =\ell_3 +m_3+2|m_2|,
\end{equation}
\begin{equation}
s_1 =\ell_2+\ell_3+m_3+2|m_2|,\quad\quad n_1 =\ell_2 +\ell_3 +m_3 +2m_1 +2|m_2|.
\end{equation}
Therefore, the expression for the angle defect along $\Gamma_2$ becomes
\begin{equation}
b_2 =\log\left[\frac{(\ell_2+2|m_2|)(\ell_2+2m_1)(\ell_2 +\ell_3 +2|m_2|)
(\ell_2 +\ell_3 +2m_1+2m_3+2|m_2|)}{\ell_2(\ell_2 +2m_1 +2|m_2|)(\ell_2+\ell_3+2m_3+2|m_2|)(\ell_2+\ell_3+2m_1+2|m_2|)}\right].
\end{equation}
Consider the special case in which $m_1=m_3$ and $\ell_2=\ell_3 =:\ell$. Then setting $\mu_1=m_1 /\ell$ and $\mu_2 =|m_2|/\ell$ produces
\begin{equation}
b_2 =\log\left[\frac{(1+2\mu_2)(1+2\mu_1)(1+\mu_2)(1+2\mu_1 +\mu_2)}
{(1+2\mu_1 +2\mu_2)(1+\mu_1 +\mu_2)^2}\right].
\end{equation}
Observe that if $\mu_1 >0$ is fixed and $\mu_2 \rightarrow 0$ then
\begin{equation}
b_2 \rightarrow \log\left[\frac{1+2\mu_1}{(1+\mu_1)^2}\right]<0.
\end{equation}
On the other hand, if $\mu_1 >0$ is fixed and $\mu_2 \rightarrow\infty$ then
\begin{equation}
b_2 \rightarrow \log(1+2\mu_1)>0.
\end{equation}
Thus, by the intermediate value theorem, there exist choices of $m_1 =m_3>0$, $m_2 <0$, and $\ell_2 =\ell_3=\ell>0$ so that the conical singularity along $\Gamma_2$ is resolved, $b_2 =0$. By symmetry, we also have that $b_3=0$, and therefore the resulting static vacuum spacetime is regular outside of the negative mass singularities.

In the next subsection, the ADM mass of static vacuum solutions obtained by superposition of signed masses will be computed. From this we find that the ADM mass of the balanced solution above is positive precisely when $2\mu_1 >\mu_2$. To confirm that this is indeed the case, it is sufficient to observe that $b_2>0$ when $2\mu_1 =\mu_2$ and $\mu_1>0$.

In a similar manner, it may be verified that there is a configuration with a reflection symmetry, consisting of two negative mass singularities having Komar masses $m_1, m_3<0$ and a single horizon of mass $m_2>0$, that is devoid of conical singularities along the axes. Moreover, this solution has negative ADM mass.

\subsection{The ADM mass of superpositioned signed masses in 4 dimensions}
\label{massec1}

Let $\xi$ denote the dual 1-form to the Killing field $\partial_t$. Then the ADM mass may be computed via a Komar integral as
\begin{equation}\label{ADM-komar-4d}
m= -\frac{1}{8\pi} \int_{S_\infty^2} \star d\xi,
\end{equation}
where $S_\infty^{2}$ denotes a limit of coordinate spheres $S_r^2$ in the asymptotically flat end as $r\rightarrow\infty$. Write $u=2\log\rho +\bar{u}$ and observe that
\begin{equation}
g = - e^{-\bar u}  dt^2 + e^{\bar u} \rho^2 d \phi^2 + e^{2 \alpha} (d \rho^2+ dz^2).
\end{equation}
We then find that
\begin{equation}\label{polarform}
\star d \xi = \rho  e^{2 \alpha - \bar{u}} (- \bar{u}_\rho dz \wedge d\phi + \bar{u}_z d \rho \wedge d \phi)=\bar{u}_r e^{2\alpha -\bar{u}} r^2 \sin\theta d\theta \wedge d\phi,
\end{equation}
where polar coordinates defined by $\rho=r\sin\theta$, $z=r\cos\theta$, have been used.
From \eqref{einsteineq}, the quadrature equations for $\alpha$ may be expressed in terms of $\bar{u}$ by
\begin{equation}\label{alphau}
\left(\alpha-\frac{1}{2}\bar{u}\right)_{\rho}=\frac{\rho}{4}\left(\bar{u}_{\rho}^2 -\bar{u}_{z}^2\right),\quad\quad\quad \left(\alpha-\frac{1}{2}\bar{u}\right)_{z}
=\frac{\rho}{2}\bar{u}_{\rho} \bar{u}_{z}.
\end{equation}
Furthermore, since $\bar{u}$ is harmonic an expansion in spherical harmonics in the asymptotically flat end yields
\begin{equation}
\bar{u}=\frac{c}{r}+O(r^{-2}),
\end{equation}
for some constant $c$. It then follows from \eqref{alphau} that $|\nabla(2\alpha-\bar{u})|=O(r^{-1})$, and therefore \eqref{polarform} implies that
\begin{equation}
\star d\xi=\left(-\frac{c}{r^2}+O(r^{-3})\right)\left(1+O(r^{-1})\right)r^2 \sin\theta d\theta\wedge d\phi.
\end{equation}
We may now evaluate the Komar mass integral to find $m=c/2$, or rather
\begin{equation}\label{massubar}
\bar{u}=\frac{2m}{r}+O(r^{-2}),
\end{equation}
as expected.

Consider now a general asymptotically flat rod structure with horizon rods $\Gamma_{h}^i$, $i=1,\ldots, i_0$ having masses $m_i>0$, and NMS rods $\Gamma_s^j$, $j=1,\ldots,j_0$ having masses $m_j <0$, then
\begin{equation}\label{baruu}
\bar{u}=-\sum_{i=1}^{i_0} G_{\Gamma_h^i} +\sum_{j=1}^{j_0} G_{\Gamma_s^j}.
\end{equation}
Suppose that the $z$-components of the north and south poles of the horizon rods are labelled $n_i$, $s_i$, and that the $z$-components of the NMS rods are labelled $n_j$, $s_j$. Then a calculation shows that the expansion of the Green's functions at infinity takes the form
\begin{equation}
G_{\Gamma_h^i}=\frac{s_i - n_i}{r}+O(r^{-2}),\quad\quad\quad
G_{\Gamma_s^j}=\frac{s_j -n_j}{r}+O(r^{-2}).
\end{equation}
We have thus obtained the following formula for the ADM mass.

\begin{lemma}
Consider a 4-dimensional, asymptotically flat, axisymmetric, static vacuum configuration consisting of the superposition of Schwarzschild solutions having masses $m_i>0$, $i=1,\ldots, i_0$, and Schwarzschild negative mass singularities having masses $m_j<0$, $j=1,\ldots, j_0$. If the corresponding horizon and NMS rods have $z$-components in Weyl-Papapetrou coordinates given by $n_i$, $s_i$, $n_j$, $s_j$ respectively, then the ADM mass of this configuration is
\begin{equation}
m=\frac{1}{2}\left(\sum_{i=1}^{i_0}(n_i -s_i)-\sum_{j=1}^{j_0}(n_j -s_j)\right)
=\sum_{i=1}^{i_0}m_i +\sum_{j=1}^{j_0}m_j.
\end{equation}
\end{lemma}

\section{5D Background and Negative Mass Black Holes}
\label{sec5} \setcounter{equation}{0}
\setcounter{section}{5}

Consider the domain of outer communication $\mathcal{M}^5$
of a static bi-axisymmetric 5-dimensional spacetime. Under mild hypotheses \cite{HollandsYazadjiev}, the orbit space $\mathcal{M}^5/[\mathbb{R}\times U(1)^2]$ is homeomorphic to the right half plane $\{(\rho,z)\mid \rho>0\}$. Assuming the ansatz that the generators of the $U(1)^2$ symmetry are orthogonal, the spacetime metric may then be expressed in Weyl-Papapetrou coordinates by
\begin{equation}\label{metricg1}
g = - e^{-u-v}\rho^2 dt^2 + e^u  d \phi^2 +e^v d\psi^2 +  e^{2 \alpha} (d \rho^2 + dz^2),
\end{equation}
where $\partial_{\phi}$, $\partial_{\psi}$ are the generators of the rotational symmetries with $\phi,\psi\in[0,2\pi)$. The vacuum Einstein equations in this setting \cite{Ma,Harmark} reduce to
the Laplace equation for $u$ and $v$, and a set of quadrature equations for $\alpha$, on $\mathbb{R}^3\setminus \{z-\text{axis}\}$ parameterized by the cylindrical coordinates $(\rho,z,\varphi)$, namely
\begin{align}\label{ealp}
\begin{split}
\Delta u=0, \quad \quad &\quad \quad\alpha_\rho =
\frac\rho4 \left[ u_\rho^2 - u_z^2 + v_\rho^2 - v_z^2 + u_\rho v_\rho - u_z v_z - \frac2\rho (u_\rho+v_\rho) \right], \\
\Delta v=0, \quad\quad &\quad\quad \alpha_z = \frac\rho4 \left[ 2u_\rho u_z + 2v_\rho v_z + u_\rho v_z + u_z v_\rho - \frac2\rho(u_z+v_z) \right].
\end{split}
\end{align}
Notice that the integrability conditions for the $\alpha$ equations correspond to the harmonicity of $u$ and $v$. Furthermore, as in the 4-dimensional case, the $z$-axis is decomposed into an exhaustive sequence of intervals called \textit{rods}, that are denoted by $\{\Gamma_l\}_{l\in I}$ for some index set $I$.
Traditionally, rods come in two types, those on which $|\partial_t|$ vanishes are referred to as \textit{horizon rods}, and those on which an integral linear combination $p\partial_{\phi}+q\partial_{\psi}$ vanishes
are referred to as \textit{axis rods} with the vector $(p,q)$ denoting the associated \textit{rod structure}. In the current setting, the possible rod structures are simply $(1,0)$ or $(0,1)$, which describe rods where $\partial_{\phi}$ or $\partial_{\psi}$ vanish. Due to this simplified rod structure profile, the only possible horizon cross-section topologies are the sphere $S^3$ and ring $S^1 \times S^2$, arising respectively when a horizon rod is bordered by $(1,0)$ and $(0,1)$ axis rods, and when a horizon rod is bordered by two $(1,0)$ axis rods or two $(0,1)$ axis rods. A topological classification of the domains of outer communication may be found in \cite{KMWY}.
The intersection point of two axis rods is called a \textit{corner}, while the intersection point of an axis rod with a horizon rod is called a \textit{pole}. Below we will introduce two new types of rods, in the 5-dimensional regime, based on negative mass singularities.

According to the above description, any solution of the bi-axisymmetric static vacuum Einstein equations, with orthogonal $U(1)$ generators, may be obtained by specifying two harmonic functions on $\mathbb{R}^3$ that have appropriate blow-up behavior on portions of the $z$-axis. As in 4-dimensions, the resulting spacetime metric may have conical singularities that form on the axis rods, and these singularities may be measured by their \textit{logarithmic angle defects}.
For an axis rod $\Gamma_1$ having the rod structure $(1,0)$, and an axis rod $\Gamma_2$ having the rod structure $(0,1)$, the corresponding logarithmic angle defects may be computed with the help of \eqref{csingularity} and are given by
\begin{equation}\label{angdef}
b_1 =  \lim_{\rho \rightarrow 0}  \left(\log \rho +\alpha-\frac{1}{2}u\right)\quad\text{ on }\quad\Gamma_1, \quad \quad\quad  b_2 = \lim_{\rho \rightarrow 0} \left(\log \rho+\alpha-\frac{1}{2}v\right)\quad\text{ on }\quad\Gamma_2 .
\end{equation}
It follows from \eqref{ealp} that these quantities are constant along their associated axis rods.  In the absence of a conical singularity, or rather a zero logarithmic angle defect, the metric is smoothly extendable across the axis. Conical singularities along an axis rod $\Gamma_l$ are classified as an \textit{angle deficit} if $b_l>0$, and an \textit{angle surplus} if $b_l<0$. Although the sign of the logarithmic angle defect determines the character of the force associated with the axis rod as in the 4-dimensional setting, unlike the lower dimensional case the force is in general not constant along axis rods. This fact will be established in Section \ref{force}.

\subsection{The signed mass Schwarzschild-Tangherlini solution}
\label{sts}

Here we will derive the Weyl-Papapetrou coordinate presentation of the negative mass Schwarzschild-Tangherlini solution. First, however, we recall the corresponding presentation for the positive mass solution. In this case, the domain of outer communication has topology $\mathbb{R}\times \left(\mathbb{R}^4 \setminus \text{Ball}\right)$, and in Schwarzschild coordinates the spacetime metric takes the form
\begin{equation}\label{stmetric}
g_{ST}= - \left( 1 - \frac{\mu}{\bar{r}^2} \right) dt^2 + \left( 1 - \frac{\mu}{\bar{r}^2} \right)^{-1} d\bar{r}^2 + \bar{r}^2 (d \bar{\theta}^2 + \sin^2 \bar{\theta} d \phi^2 +\cos^2 \bar{\theta} d\psi^2),
\end{equation}
where $\mu=\tfrac{4m}{3\pi}$ with the mass $m>0$, and $\bar{r}\geq \sqrt{\mu}$. The portion of the metric involving the angular variables, $0\leq\bar{\theta}\leq \pi/2$ and $0\leq\phi,\psi<2\pi$, represents the round metric on $S^3$ in Hopf coordinates. As in 4-dimensions, the cylindrical radius $\rho$ of Weyl-Papapetrou coordinates is obtained from the determinant of the Killing $t\phi\psi$-portion of the metric. The $z$-coordinate is then chosen as a harmonic conjugate, so that $(\rho,z)$ form isothermal coordinates on the orbit space $\mathcal{M}^5 /[\mathbb{R}\times U(1)^2]$. It follows that (see \cite[(5.13)]{Harmark})
\begin{equation}
\rho = \bar{r}^2 \sin (2\bar{\theta})  \sqrt{1 - \tfrac{\mu}{\bar{r}^2}} ,\quad\quad\quad
z  = \left(\bar{r}^2-\frac{\mu}{2}\right) \cos (2\bar{\theta}) .
\end{equation}
The spacetime metric $g_{ST}$ then takes the form \eqref{metricg1}, where
\begin{equation}
u=\log\left(\frac{\rho^2}{4(\bar{r}^2 -\mu)\cos^2 \bar{\theta}}\right)
=\log\left(\frac{\tfrac{1}{2}\rho^2}{\sqrt{\rho^2 +(z-\mu/2)^2}+(z-\mu/2)}\right)
=G_{\{z>\mu/2\}}-\log 2,
\end{equation}
\begin{equation}
v=\log\left(\frac{\rho^2}{4(\bar{r}^2 -\mu)\sin^2 \bar{\theta}}\right)
=\log\left(\frac{\tfrac{1}{2}\rho^2}{\sqrt{\rho^2 +(z+\mu/4)^2}-(z+\mu/4)}\right)
=G_{\{z<-\mu/2\}}-\log 2.
\end{equation}
From this we find that the $z$-axis is broken into two axis rods $\Gamma_{-}=(-\infty,-\mu/2)$ and $\Gamma_{+}=(\mu/2,\infty)$
having rod structures $(0,1)$ and $(1,0)$ respectively, and one horizon rod $\Gamma_{h}=(-\mu/2,\mu/2)$ which corresponds to the surface $\bar{r}=\sqrt{\mu}$ in Schwarzschild coordinates.

Consider now the negative mass Schwarzschild-Tangherlini solution. The topology of the domain of outer communication is the same as in the positive mass case, although geometrically it may be preferable to think of a point being removed from $\mathbb{R}^4$, instead of a ball, since the singularity has zero area. The metric expression \eqref{stmetric} still remains valid, although here the mass parameter $m<0$ and the areal radius extends to the origin $\bar{r}>0$. Notice that there is no longer a coordinate singularity at $\bar{r}=\sqrt{|\mu|}$, and there is no event horizon so that $\bar{r}=0$ is a naked singularity. The formulas for the change to Weyl-Papapetrou coordinates have the same form
\begin{equation}
\rho = \bar{r}^2 \sin (2\bar{\theta})  \sqrt{1 + \tfrac{|\mu|}{\bar{r}^2}} ,\quad\quad\quad
z  = \left(\bar{r}^2+\frac{|\mu|}{2}\right) \cos (2\bar{\theta}) .
\end{equation}
as do the metric coefficient harmonic functions
\begin{equation}
u=G_{\{z>-|\mu|/2\}},\quad\quad\quad\quad v=G_{\{z<|\mu|/2\}}.
\end{equation}
There is, however, a change in the rod structure which will require the introduction of a new type of rod.

\begin{figure}
\includegraphics[width=.9\textwidth]{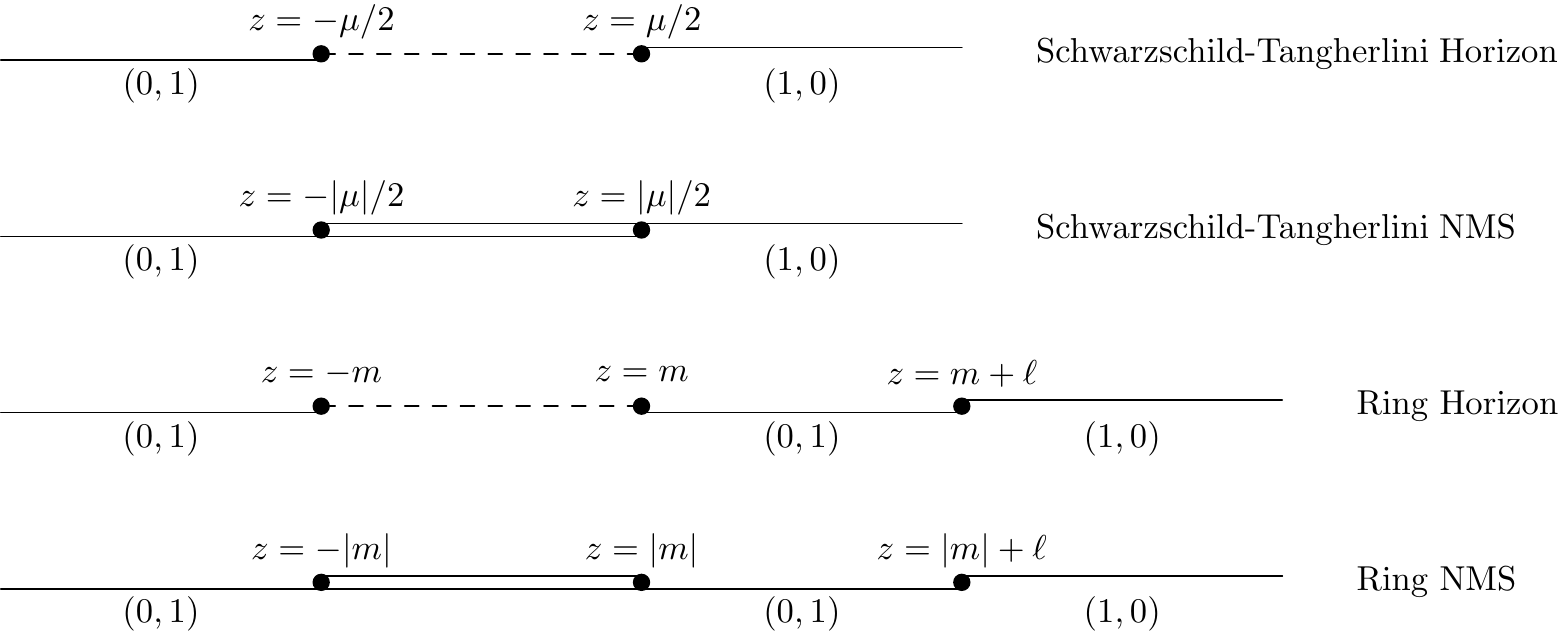}
\caption{Rod structures for Schwarzschild-Tangherlini and static ring metrics of positive and negative mass. The dotted lines indicate a horizon rod, a single solid line indicates an axis rod, while overlapping solid lines indicate an NMS rod.}
\label{5d-schw-zas1}
\end{figure}

The rod structure is determined by the asymptotic behavior of $u$ and $v$ upon approach to the $z$-axis. Observe that the singular support sets for the Green's functions that comprise $u$ and $v$, overlap on the interval $\Gamma_s =(-|\mu|/2,|\mu|/2)$. As above, let $\Gamma_{-}=(-\infty,-|\mu|/2)$ and $\Gamma_{+}=(|\mu|/2,\infty)$, then it follows from \eqref{ealp} that as $\rho\rightarrow 0$ we have
\begin{equation}
u=
\begin{cases}
2\log\rho +O(1) & \text{if $z\in \Gamma_s \cup\Gamma_{+}$} \\
O(1) & \text{if $z\in\Gamma_{-}$}
\end{cases},
\quad\quad\quad
v=
\begin{cases}
2\log\rho +O(1) & \text{if $z\in \Gamma_s \cup\Gamma_{-}$} \\
O(1) & \text{if $z\in\Gamma_{+}$}
\end{cases},
\end{equation}
\begin{equation}
\alpha=
\begin{cases}
O(1) & \text{if $z\in \Gamma_{\pm}$} \\
\log\rho +O(1) & \text{if $z\in\Gamma_s$}
\end{cases}.
\end{equation}
This shows that the rods $\Gamma_{-}$, $\Gamma_{+}$ behave as typical axis rods having rod structures $(0,1)$, $(1,0)$ respectively, whereas $\Gamma_s$ does not exhibit the characteristic of either an axis or horizon rod.
We will therefore refer to $\Gamma_s$, and more generally any rod of an bi-axisymmetric static spacetime near which $u=2\log\rho +O(1)$ and $v=2\log\rho +O(1)$, as a \textit{spherical negative mass singularity rod}. These intervals of the $z$-axis correspond to single points in time slices of the spacetime, and may be viewed as the overlap of two neighboring axis rods having different rod structures, see Figure \ref{5d-schw-zas1}. In this way negative mass singularities in 5-dimensions can be thought of as spread out corners. More precisely, the overlap of the singular support is a point for a corner, whereas the overlap is an interval for a NMS.

\subsection{The signed mass static black ring solution}
\label{sbr}

Consider the rod configuration for an asymptotically flat positive mass static black hole with ring $S^1\times S^2$ horizon cross-section topology. Namely, there are two semi-infinite axis rods $\Gamma_1=(m+l,\infty)$ and $\Gamma_4 =(-\infty,-m)$ having rod structures $(1,0)$ and $(0,1)$ respectively, where the parameters $m$ and $l$ are positive. In addition, there is a finite axis rod $\Gamma_2=(m,m+l)$ having rod structure $(0,1)$, and a horizon rod $\Gamma_3 =(-m,m)$, as well as a corner point at $z=m+l$. Note that the parameter $l$ represents the length of the finite axis rod, and $\tfrac{3\pi}{2} m$ is the ADM mass. The relevant spacetime metric \eqref{metricg1} may be constructed by setting
\begin{equation}
u=G_{\{z>m+l\}},\quad\quad\quad\quad v=G_{\{z<-m\}}+G_{\{m<z<m+l\}},
\end{equation}
and solving for $\alpha$ from \eqref{ealp}. This solution possesses a conical singularity along the finite axis rod $\Gamma_2$ for any choice of parameters. Positive mass static black ring solutions were studied in detail in \cite{EmparanReall}.

Here we introduce the negative mass static ring solution. This is obtained from the above configuration by letting $m<0$, replacing $m$ with $|m|$ in the rod intervals $\Gamma_i$, $i=1,2,3,4$, and changing $\Gamma_3$ to a \textit{ring negative mass singularity rod}. This last step is achieved, with motivation from the spherical NMS example of the previous subsection, by requiring both harmonic functions $u$ and $v$ to have asymptotics $2\log\rho+O(1)$ upon approach to this rod. Thus, the desired solution may be constructed from
\begin{equation}
u=G_{\{z>|m|+l\}}+G_{\{-|m|<z<|m|\}},\quad\quad\quad\quad v=G_{\{z<|m|+l\}}.
\end{equation}
As shown in Section \ref{massec1}, the ADM mass of this solution is $\tfrac{3\pi}{2} m<0$. Furthermore, from the analysis of angle defects in the next section, we find that this solution also possesses a conical singularity on the finite axis rod for any choice of parameters. Geometrically, a neighborhood of the NMS singularity (within a time slice) may be interpreted as a cone with $S^1\times S^2$ cross-sections where the ring NMS rod represents the cone vertex.

Trivial examples having negative mass that are free of conical singularities may be obtained, although they will not be asymptotically flat. In particular, 5-dimensional static black ring solutions of signed mass may be constructed from the 4-dimensional Schwarzschild solutions of signed mass by taking the product with a circle of fixed length. If $\psi$ is the coordinate parameterizing the circle of fixed length, then  $v=1$, and in the negative mass case $u=4\log \rho +O(1)$ as $\rho\rightarrow 0$ at the horizon. The negative mass singularity here is topologically a circle, as opposed to a point, and the solution is asymptotically Kaluza-Klein. We also mention that these circle type negative mass singularities may superpositioned together with horizons, as well as the point type negative mass singularities, to form more complicated rod structures.

\subsection{The force of axis struts in 5D}
\label{force}

Following the exposition of \cite[Section 5]{Weinstein1990}, we will compute the force along an axis rod induced by a conical singularity. This calculation is valid
for 5-dimensional static (and more generally stationary) bi-axisymmetric spacetimes, and differs from the 4-dimensional case treated in \cite{Weinstein1990} in that that force is not necessarily constant along the axis.

Let $p$ be an interior point of an axis rod $\Gamma$, which we may assume without loss of generality has rod structure $(1,0)$, and denote the logarithmic angle defect by $b$ as in \eqref{angdef}. The cone angle is encoded in the disc $D_{\rho}$ of (coordinate) radius $\rho$, which passing through $p$, and is obtained by fixing values for $(t,z,\psi)$. The induced metric is
\begin{equation}
ds^2 =e^{2\alpha}d\rho^2 +e^{u} d\phi^2,
\end{equation}
and the geodesic curvature of the boundary $\partial D_{\rho}$ takes the form
$\kappa=\tfrac{1}{2}e^{-\alpha}u_{\rho}$. It follows that
\begin{equation}\label{7uyt68}
\int_{\partial D_{\rho}}\kappa=\pi e^{u/2 -\alpha} u_{\rho}=\pi e^{\bar{u}/2-\alpha}
(2+\rho\bar{u}_{\rho}),
\end{equation}
where $\bar{u}=u-2\log\rho$ is a smooth function up to the axis. If a conical singularity is present, the Gauss curvature of the disc $D_{\rho}$ is singular at the origin. Nevertheless, the Gauss curvature defines a signed measure $\mu_{K}$, and in light of the Gauss-Bonnet Theorem along with \eqref{7uyt68} we have
\begin{equation}\label{llll}
\lim_{\rho\rightarrow 0} \mu_{K}\left(D_{\rho}\right)
=\lim_{\rho\rightarrow 0}\left(2\pi -\int_{\partial D_{\rho}}\kappa\right)
=2\pi\left(1-e^{-b}\right).
\end{equation}

Let $\{e_i\}_{i=0}^{4}$ be an orthonormal frame such that $e_0$ is timelike, $e_1$ and $e_2$ are multiples of $\partial_{z}$ and $\partial_{\rho}$, and $e_3$, $e_4$ are tangent to the toroidal fibers, with $e_2$ and $e_3$ spanning that tangent space to $D_{\rho}$. Observe that the only singular sectional curvature is that associated with the $e_2 e_3$-tangent plane. In particular, according to the Gauss equations
\begin{equation}
R_{2323}=K+\textrm{bounded terms},
\end{equation}
where the bounded terms are quadratic expressions in the second fundamental form of the disc $D_{\rho}$. This boundedness follows from the fact that the disc
intersects the $z$-axis perpendicularly at $p$, and hence the second fundamental form, which measures the `turning of the unit normal', remains finite. It follows that the only singular Ricci curvatures are $R_{22}$ and $R_{33}$, both of which agree with $K$ up to bounded expressions. Therefore the Einstein equations (expressed in geometrized units) yield
\begin{equation}\label{pppp}
8\pi T_{11}=R_{11}-\frac{1}{2}R g_{11}=-K+\textrm{bounded terms}.
\end{equation}
This indicates the presence of `matter' along an axis rod having a nonzero angle defect. Since the stress-energy-momentum tensor component $T_{11}$ measures normal stress (in the $z$-direction) or rather force per unit volume, the force at $p$ arising from this `matter distribution' is given by
\begin{equation}
F(p)=\lim_{\rho\rightarrow0}\int_{0}^{2\pi}\mu_{T_{11}}(D_{\rho}) e^{v/2}d\psi
=\frac{\pi e^{v(p)/2}}{2}\left(e^{-b}-1\right),
\end{equation}
where in the second equality we have used \eqref{llll} and \eqref{pppp}. Comparing with the 4-dimensional case \cite[(36)]{Weinstein1990}, we find that although it still holds that an angle surplus (deficit) produces a positive (negative) force, the force is no longer necessarily constant along the axis rod due to the presence of the new term $e^{v/2}$.

\section{Conical Singularities in Dimension 5}
\label{sec6} \setcounter{equation}{0}
\setcounter{section}{6}

The goal of this section is to analyze the propagation of conical singularities across spherical and ring horizons and NMSs, as well as corners, for bi-axisymmetric static vacuum solutions. Formulas for the difference of the logarithmic angle defect between two neighboring axis rods will be given in terms of renormalized values for $u$ and $v$ at the poles, or at a corner. Let $z_2 < z_s <z_n < z_1$ denote values on the $z$-axis, and consider a sequence of two axis rods $\Gamma_1 =(z_n , z_1)$ to the north having rod structure $(1,0)$, and $\Gamma_2 =(z_2, z_s)$ to the south, that border a horizon/NMS rod $\Gamma_{h/s} =(z_s, z_n)$ with poles $N$, $S$ located at $z_n$, $z_s$. Since the angle defects are constant along the axes, the relation between the two logarithmic angle defects may be computed at any two points along these two axes. Namely, let $\varepsilon>0$ be a small parameter and use points $N_{\varepsilon}=(0,z_n +\varepsilon)\in \Gamma_1$ and $S_{-\varepsilon}=(0,z_s -\varepsilon)\in \Gamma_2$ to find that for a spherical horizon/NMS in which $\Gamma_2$ has rod structure $(0,1)$, formula \eqref{angdef} implies
\begin{align}\label{mainform1}
\begin{split}
b_1 -b_2 =& \lim_{\rho\rightarrow 0}\left(\alpha-\frac{1}{2}u+\log\rho\right)\Big|_{z=z_n +\varepsilon} -\lim_{\rho\rightarrow 0}\left(\alpha-\frac{1}{2}v+\log\rho\right)\Big|_{z=z_s - \varepsilon}\\
=& \frac{1}{2}\left(\bar{v}(S_{-\varepsilon})-\bar{u}(N_{\varepsilon})\right)
+\alpha(N_{\varepsilon})-\alpha(S_{-\varepsilon}),
\end{split}
\end{align}
where $\bar{u}=u-2\log\rho$ and $\bar{v}=v-2\log\rho$. Similarly, for a ring horizon/NMS in which $\Gamma_2$ has rod structure $(1,0)$ we have
\begin{align}\label{mainform2}
\begin{split}
b_1 -b_2 =& \lim_{\rho\rightarrow 0}\left(\alpha-\frac{1}{2}u+\log\rho\right)\Big|_{z=z_n +\varepsilon} -\lim_{\rho\rightarrow 0}\left(\alpha-\frac{1}{2}u+\log\rho\right)\Big|_{z=z_s - \varepsilon}\\
=& \frac{1}{2}\left(\bar{u}(S_{-\varepsilon})-\bar{u}(N_{\varepsilon})\right)
+\alpha(N_{\varepsilon})-\alpha(S_{-\varepsilon}).
\end{split}
\end{align}
In the case of a corner point at $C=(0,z_c)$, let $z_2 < z_c < z_1$ and consider two axis rods $\Gamma_1 =(z_c , z_1)$ of rod structure $(1,0)$, and $\Gamma_2 =(z_2, z_c)$ of rod structure $(0,1)$. If $C_{\varepsilon}=(0,z_c +\varepsilon)\in \Gamma_1$ and $C_{-\varepsilon}=(0,z_c -\varepsilon)\in \Gamma_2$, then
\begin{align}\label{mainform3}
\begin{split}
b_1 -b_2 =& \lim_{\rho\rightarrow 0}\left(\alpha-\frac{1}{2}u+\log\rho\right)\Big|_{z=z_c +\varepsilon} -\lim_{\rho\rightarrow 0}\left(\alpha-\frac{1}{2}v+\log\rho\right)\Big|_{z=z_c - \varepsilon}\\
=& \frac{1}{2}\left(\bar{v}(C_{-\varepsilon})-\bar{u}(C_{\varepsilon})\right)
+\alpha(C_{\varepsilon})-\alpha(C_{-\varepsilon}).
\end{split}
\end{align}
In the remainder of this section we will evaluate these expressions by taking $\varepsilon\rightarrow 0$.

\subsection{Across a corner}

In order to evaluate the right-hand side of \eqref{mainform3}, expansions for $u$, $v$, and $\alpha$ at the corner are needed to isolate the singular parts. From the Green's function analysis of previous sections, we find that
\begin{equation}\label{bv}
u=\log\left( r_c -(z-z_c)\right)+\hat{u},\quad\quad
v=\log\left( r_c +(z-z_c)\right)+\hat{v}\quad \text{ near } C,
\end{equation}
for some smooth functions $\hat{u}$ and $\hat{v}$, where $r_{c}=\sqrt{\rho^2 +(z-z_{c})^2}$ is the Euclidean distance to $C$. It follows that
\begin{equation}
\bar{u}=u-2\log\rho=\hat{u}-\log(2|z-z_c|)+O(\rho^2)\quad \text{ for } z>z_c,
\end{equation}
\begin{equation}
\bar{v}=v-2\log\rho=\hat{v}-\log(2|z-z_c|)+O(\rho^2)\quad \text{ for } z<z_c.
\end{equation}
In particular
\begin{equation}
\frac{1}{2}\left(\bar{v}(C_{-\varepsilon})-\bar{u}(C_{\varepsilon})\right)
=\frac{1}{2}\left(\hat{v}(C_{-\varepsilon})-\hat{u}(C_{\varepsilon})\right)
\rightarrow \frac{1}{2}\left(\hat{v}(C)-\hat{u}(C)\right),
\end{equation}
as $\varepsilon\rightarrow 0$. Furthermore according to \cite[(5.8)]{KhuriWeinsteinYamadaJHEP}, the expansion for $\alpha$ is found to be
\begin{equation}
\alpha=-\frac{1}{2}\log r_c +\frac{r_c +(z-z_c)}{4r_c}\left(\hat{u} -\hat{u}(C)\right)+\frac{r_c -(z-z_c)}{4r_c}\left(\hat{v} -\hat{v}(C)\right)
+\hat{c} +O(\rho^2)\quad\text{ near }C,
\end{equation}
for some constant $\hat{c}$. Therefore
\begin{equation}
\alpha(C_{\varepsilon})-\alpha(C_{-\varepsilon})
=\frac{1}{2}\left(\hat{u}(C_{\varepsilon})-\hat{u}(C)\right)
-\frac{1}{2}\left(\hat{v}(C_{\varepsilon})-\hat{v}(C)\right)
\rightarrow 0
\end{equation}
as $\varepsilon\rightarrow 0$. These observations combine to yield the desired formula for the propagation of cone angle defect across a corner.

\begin{prop}\label{6.1}
Consider a static bi-axisymmetric solution \eqref{stmetric} of the 5D vacuum Einstein equations. Let $C$ be a corner point at the intersection of two axis rods $\Gamma_1$ and $\Gamma_2$ having rod structures $(1,0)$ and $(0,1)$, respectively. Then the difference of logarithmic angle defects is given by
\begin{equation}\label{cornerad}
b_1 -b_2 =\frac{1}{2}\left(\hat{v}(C)-\hat{u}(C)\right).
\end{equation}
\end{prop}

\subsection{Across a spherical horizon}\label{ash}

Here we will compute \eqref{mainform1} for a horizon. Note that $u$ and $v$ have the following expansions in the vicinity of the north and south poles of the horizon
\begin{equation}
u=\log\left(r_n -(z-z_n)\right) +\hat{u}_n \quad \text{ near } N, \quad\quad
v=\log\left(r_s +(z-z_s)\right) +\hat{v}_s \quad \text{ near } S,
\end{equation}
for some smooth functions $\hat{u}_n$, $\hat{v}_s$ and where $r_{n/s}=\sqrt{\rho^2 +(z-z_{n/s})^2}$ is the Euclidean distance to the poles. Furthermore, observe that as in \eqref{3.5} we have
\begin{equation}\label{3.5.0}
\bar{u}(N_{\varepsilon})=\hat{u}_n (N_{\varepsilon}) -\log(2\varepsilon),\quad\quad\quad
\bar{v}(S_{-\varepsilon})=\hat{v}_s (S_{-\varepsilon})-\log(2\varepsilon).
\end{equation}
It follows that
\begin{equation}\label{mmmmm}
\frac{1}{2}\left(\bar{v}(S_{-\varepsilon})-\bar{u}(N_{\varepsilon})\right)
=\frac{1}{2}\left(\hat{v}_s(S)-\hat{u}_n(N)\right)+O(\varepsilon).
\end{equation}
The expansion for $\alpha$ at the poles is provided by \cite[(5.15)]{KhuriWeinsteinYamadaJHEP}, namely
\begin{equation}\label{1haudh}
\alpha=-\frac{1}{2}\log r_n +\frac{(z-z_n)}{2r_n}\left(\hat{u}_n -\hat{u}_n(N)\right)+\frac{[(z-z_n)-r_n]}{4r_n}\left(v-v(N)\right)
+\hat{c}_n +O(\rho^2)\quad\quad\text{ near } N,
\end{equation}
and
\begin{equation}
\alpha=-\frac{1}{2}\log r_s -\frac{(z-z_s)}{2r_s}\left(\hat{v}_s -\hat{v}_s(S)\right)-\frac{[(z-z_s)+r_s]}{4r_s}\left(u-u(S)\right)
+\hat{c}_s +O(\rho^2)\quad\quad\text{ near } S,
\end{equation}
for some constants $\hat{c}_n$, $\hat{c}_s$. We then have
\begin{equation}\label{al1}
\alpha(N_{\varepsilon})-\alpha(N_{-\varepsilon})
=\frac{1}{2}\left(\hat{u}_n(N_{\varepsilon})-\hat{u}_n(N)\right)
+\frac{1}{2}\left(\hat{u}_n(N_{-\varepsilon})-\hat{u}_n(N)\right)
+\frac{1}{2}\left(v(N_{-\varepsilon})-v(N)\right)
=O(\varepsilon),
\end{equation}
and similarly
\begin{equation}\label{al2}
\alpha(S_{\varepsilon})-\alpha(S_{-\varepsilon})=O(\varepsilon).
\end{equation}
Therefore
\begin{align}\label{al3}
\begin{split}
\alpha(N_{\varepsilon})-\alpha(S_{-\varepsilon})
=&\left[\alpha(N_{\varepsilon})-\alpha(N_{-\varepsilon})\right]
+\left[\alpha(N_{-\varepsilon})-\alpha(S_{\varepsilon})\right]
+\left[\alpha(S_{\varepsilon})-\alpha(S_{-\varepsilon})\right]\\
=&\alpha(N_{-\varepsilon})-\alpha(S_{\varepsilon})+O(\varepsilon).
\end{split}
\end{align}
Furthermore, since $u$ and $v$ are regular on the horizon, \eqref{ealp} implies that
\begin{align}\label{al4}
\begin{split}
\alpha(N_{-\varepsilon})-\alpha(S_{\varepsilon})=&\int_{z_s +\varepsilon}^{z_n -\varepsilon}\alpha_{z}(0,z) dz\\
=&-\frac{1}{2}\int_{z_s +\varepsilon}^{z_n -\varepsilon}\left(u+v\right)_{z}(0,z) dz\\
=&-\frac{1}{2}(u+v)(N_{-\varepsilon})+\frac{1}{2}(u+v)(S_{\varepsilon})\\
=&-\frac{1}{2}(\hat{u}_n+v)(N_{-\varepsilon})
+\frac{1}{2}(u+\hat{v}_s)(S_{\varepsilon})\\
=&-\frac{1}{2}(\hat{u}_n+v)(N)
+\frac{1}{2}(u+\hat{v}_s)(S)+O(\varepsilon).
\end{split}
\end{align}
Putting all this together, and taking $\varepsilon\rightarrow 0$, yields the following result.

\begin{prop}\label{6.2}
Consider a static bi-axisymmetric solution \eqref{stmetric} of the 5D vacuum Einstein equations. Let $N$ and $S$ be the poles of a spherical horizon rod that lies between two axis rods, $\Gamma_1$ and $\Gamma_2$, having rod structures $(1,0)$ and $(0,1)$ respectively. Then the difference of logarithmic angle defects is given by
\begin{equation}\label{shorizon}
b_1 -b_2=\hat{v}_s(S)-\hat{u}_n(N)+\frac{1}{2}\left(u(S)-v(N)\right).
\end{equation}
\end{prop}

\subsection{Across a ring horizon}

Consider now \eqref{mainform2} for a horizon. The expansion of $u$ in the vicinity of the north and south poles is given by
\begin{equation}
u=\log\left(r_n -(z-z_n)\right) +\hat{u}_n \quad \text{ near } N, \quad\quad
u=\log\left(r_s +(z-z_s)\right) +\hat{u}_s \quad \text{ near } S,
\end{equation}
for some smooth functions $\hat{u}_n$, $\hat{u}_s$. Note that due to the ring rod structure, the function $v$ remains smooth in a neighborhood of both poles. Furthermore, observe that as in \eqref{3.5.0} we have
\begin{equation}
\bar{u}(N_{\varepsilon})=\hat{u}_n (N_{\varepsilon}) -\log(2\varepsilon),\quad\quad\quad
\bar{u}(S_{-\varepsilon})=\hat{u}_s (S_{-\varepsilon})-\log(2\varepsilon).
\end{equation}
It follows that
\begin{equation}\label{mmmmmr}
\frac{1}{2}\left(\bar{u}(S_{-\varepsilon})-\bar{u}(N_{\varepsilon})\right)
=\frac{1}{2}\left(\hat{u}_s(S)-\hat{u}_n(N)\right)+O(\varepsilon).
\end{equation}
The expansion for $\alpha$ at the poles is again provided by \cite[(5.15)]{KhuriWeinsteinYamadaJHEP}. In fact, near the north pole
it coincides with \eqref{1haudh}, while on the other side
\begin{equation}
\alpha=-\frac{1}{2}\log r_s -\frac{(z-z_s)}{2r_s}\left(\hat{u}_s -\hat{u}_s(S)\right)-\frac{[(z-z_s)+r_s]}{4r_s}\left(v-v(S)\right)
+\hat{c}_s +O(\rho^2)\quad\quad\text{ near } S,
\end{equation}
for some constant $\hat{c}_s$. We then find that \eqref{al1}-\eqref{al3} are valid, and moreover \eqref{al4} may be slightly modified to produce
\begin{equation}
\alpha(N_{-\varepsilon})-\alpha(S_{\varepsilon})
=-\frac{1}{2}(\hat{u}_n+v)(N)
+\frac{1}{2}(\hat{u}_s +v)(S)+O(\varepsilon).
\end{equation}
Taking the limit $\varepsilon\rightarrow 0$ then produces the desired formula.

\begin{prop}\label{6.3}
Consider a static bi-axisymmetric solution \eqref{stmetric} of the 5D vacuum Einstein equations. Let $N$ and $S$ be the poles of a ring horizon rod that lies between two axis rods, $\Gamma_1$ and $\Gamma_2$, both having rod structure $(1,0)$. Then the difference of logarithmic angle defects is given by
\begin{equation}\label{rhorizon}
b_1 -b_2=\hat{u}_s(S)-\hat{u}_n(N)+\frac{1}{2}\left(v(S)-v(N)\right).
\end{equation}
\end{prop}

\subsection{Across a spherical negative mass singularity}

Let us now compute \eqref{mainform1} for a NMS. In this situation both $u$ and $v$ asymptote to $2\log\rho$ upon approach to the NMS rod $\Gamma_s$. Near the poles these functions have the expansions
\begin{equation}
u=2\log\rho +\bar{u}, \quad\quad
v=\log\left(r_n +(z-z_n)\right) +\tilde{v}_n \quad \text{ near } N,
\end{equation}
and
\begin{equation}
u=\log\left(r_s -(z-z_s)\right) +\tilde{u}_s, \quad\quad
v=2\log\rho +\bar{v} \quad \text{ near } S,
\end{equation}
for some smooth functions $\bar{u}$, $\bar{v}$, $\tilde{u}_s$, and $\tilde{v}_n$ where $r_{n/s}=\sqrt{\rho^2 +(z-z_{n/s})^2}$ is the Euclidean distance to the poles. From the point of view of $u$ the north pole exhibits the character of an axis rod point, while the same is true of $v$ with regards to the south pole. Therefore the first expression on the right-hand side of \eqref{mainform1} may simply be evaluated as
\begin{equation}\label{mmmmm1}
\frac{1}{2}\left(\bar{v}(S_{-\varepsilon})-\bar{u}(N_{\varepsilon})\right)
=\frac{1}{2}\left(\bar{v}(S)-\bar{u}(N)\right)+O(\varepsilon).
\end{equation}
Evaluation of the remaining terms in \eqref{mainform1} is more difficult, as the behavior of $\alpha$ near the poles of NMSs is more complex.

To proceed with the expansion of $\alpha$, we will make use of
\begin{equation}
\partial_{\rho}\log\left(r_n +(z-z_n)\right)=\frac{\rho}{r_n [r_n +(z-z_n)]},\quad\quad \partial_{\rho}\log\left(r_s -(z-z_s)\right)=\frac{\rho}{r_s [r_s -(z-z_s)]},
\end{equation}
\begin{equation}
\partial_{z}\log\left(r_n +(z-z_n)\right)=\frac{1}{r_n},\quad \quad \partial_{z}\log\left(r_s -(z-z_s)\right)=-\frac{1}{r_s}.
\end{equation}
Using this, together with the formulas for $\alpha$ in \eqref{ealp}, produces the following expression near $N$ after a lengthy calculation
\begin{align}\label{rho1}
\begin{split}
\alpha_{\rho}=&\frac{\rho}{4 r_n^2}\left(\frac{\rho^2}{[r_n +(z-z_n)]^2}
-1\right)+\frac{1}{4}\left(2+\frac{\rho^2}{r_n[r_n +(z-z_n)]}\right)\bar{u}_{\rho}
+\frac{\rho^2}{2r_n[r_n +(z-z_n)]}(\tilde{v}_{n})_{\rho}\\
&-\frac{\rho}{4r_n}\bar{u}_z -\frac{\rho}{2r_n}(\tilde{v}_{n})_{z}+O(\rho),
\end{split}
\end{align}
and similarly
\begin{align}\label{z1}
\begin{split}
\alpha_{z}=&\frac{\rho^2}{2r_n^2[r_n +(z-z_n)]}+\frac{1}{4}\left(2+\frac{\rho^2}{r_n[r_n +(z-z_n)]}\right)\bar{u}_{z}+\frac{\rho^2}{2r_n[r_n +(z-z_n)]}(\tilde{v}_{n})_{z}\\
&+\frac{\rho}{4r_n}\bar{u}_{\rho} +\frac{\rho}{2r_n}(\tilde{v}_{n})_{\rho}+O(\rho^2).
\end{split}
\end{align}
Consider now polar coordinates centered at the north pole so that $\rho=r_n \sin\theta_n$ and $z-z_n=r_n \cos\theta_n$, and integrate the radial derivative to find the expansion
\begin{align}\label{alphatheta}
\begin{split}
\alpha=&\int_0^{r_n}\alpha_{s}(s,\theta_n) ds \\
=& \int_0^{r_n}\left(\sin\theta_n \alpha_{\rho}+\cos\theta_n \alpha_z \right)ds\\
=&\frac{1}{4}\left(2+\frac{\rho^2}{r_n[r_n +(z-z_n)]}\right)\left(\bar{u}-\bar{u}(N)\right)
+\frac{\rho^2}{2r_n[r_n +(z-z_n)]}\left(\tilde{v}_{n}-\tilde{v}_n (N)\right)
+c_n(\theta_n)+O(\rho^2)
\end{split}
\end{align}
about $N$, for some function $c_n(\theta_n)$. In order to find this function we may compute the angular derivative of $\alpha$ in two different ways. First, using \eqref{rho1} and \eqref{z1} we obtain
\begin{align}
\begin{split}
\alpha_{\theta_n}=&r_n \cos\theta_n \alpha_{\rho} -r_n \sin\theta_n \alpha_z\\
=&-\frac{\sin\theta_n}{2(1+\cos\theta_n)}+\frac{1}{4}\left(2+\frac{\sin^2 \theta_n}{1+\cos\theta_n}\right)\bar{u}_{\theta_n}
+\frac{\sin^2 \theta_n}{2(1+\cos\theta_n)}(\tilde{v}_n)_{\theta_n}\\
=&-\frac{r_n \sin\theta_n}{4}\left(\bar{u}_{r_n}+2(\tilde{v})_{r_n}\right)+O(r_n \rho^2).
\end{split}
\end{align}
On the other hand we may differentiate \eqref{alphatheta} directly, with respect to $\theta_n$, and compare the two results to find that there is a constant $\tilde{c}_n$ such that
\begin{equation}
c_n(\theta)=\frac{1}{2}\log(1+\cos\theta_n)+\tilde{c}_n.
\end{equation}
It follows that
\begin{align}\label{+++++}
\begin{split}
\alpha=&\frac{1}{2}\log(1+\cos\theta_n)+\frac{1}{4}\left(2+\frac{\rho^2}{r_n[r_n +(z-z_n)]}\right)\left(\bar{u}-\bar{u}(N)\right)\\
&+\frac{\rho^2}{2r_n[r_n +(z-z_n)]}\left(\tilde{v}_{n}-\tilde{v}_n (N)\right)
+\tilde{c}_n +O(\rho^2)\quad\quad\quad\quad\text{ near }N.
\end{split}
\end{align}
Similar manipulations may be used to show that
\begin{align}\label{-----}
\begin{split}
\alpha=&\frac{1}{2}\log(1-\cos\theta_s)+\frac{\rho^2}{2r_s[r_s -(z-z_s)]}\left(\tilde{u}_{s}-\tilde{u}_s (S)\right)\\
&+\frac{1}{4}\left(2+\frac{\rho^2}{r_s[r_s -(z-z_s)]}\right)\left(\bar{v}-\bar{v}(S)\right)
+\tilde{c}_s +O(\rho^2)\quad\quad\quad\text{ near }S,
\end{split}
\end{align}
for some constant $\tilde{c}_s$ where $\rho=r_s \sin\theta_s$ and $z-z_s=r_s \cos\theta_s$.

The computations \eqref{+++++}, \eqref{-----} demonstrate that near the interior of the NMS rod, $\alpha\sim\log\rho$ as $\rho\rightarrow 0$. In particular, in contrast to horizon rods, $\alpha$ blows-up upon approach to an NMS rod. For this reason, we cannot directly use the argument of Section \ref{ash} in which the $\alpha$ terms of \eqref{mainform1} are computed by integrating along the horizon rod. Thus, we will instead integrate along vertical segments slightly off of the NMS rod. To do this, let $\delta>0$ be a small parameter, and define points $N_{-\varepsilon}^{\delta}=(\delta,z_n -\varepsilon)$ and $S_{\varepsilon}^{\delta}=(\delta,z_s +\varepsilon)$. We then have
\begin{equation}
\alpha(N_{\varepsilon})-\alpha(N_{-\varepsilon}^{\delta})
=\frac{1}{2}\log 2 -\frac{1}{2}\log(1+\cos\theta_n(\varepsilon,\delta))+O(\varepsilon+\delta),
\end{equation}
and
\begin{equation}
\alpha(S_{\varepsilon}^{\delta})-\alpha(S_{-\varepsilon})
=-\frac{1}{2}\log 2 +\frac{1}{2}\log(1-\cos\theta_s(\varepsilon,\delta))+O(\varepsilon+\delta),
\end{equation}
where $\theta_{n/s}(\varepsilon,\delta)$ is the angular coordinate for $N_{-\varepsilon}^{\delta}$, $S_{\varepsilon}^{\delta}$ respectively.
Since $\theta_n(\varepsilon,\delta)=\pi -\theta_s(\varepsilon,\delta)$ it follows that
\begin{align}\label{,./}
\begin{split}
\alpha(N_{\varepsilon})-\alpha(S_{-\varepsilon})=&
\left[\alpha(N_{\varepsilon})-\alpha(N_{-\varepsilon}^{\delta})\right]
+\left[\alpha(N_{-\varepsilon}^{\delta})-\alpha(S_{\varepsilon}^{\delta})\right]
+\left[\alpha(S_{\varepsilon}^{\delta})-\alpha(S_{-\varepsilon})\right]\\
=&\alpha(N_{-\varepsilon}^{\delta})-\alpha(S_{\varepsilon}^{\delta})
+O(\varepsilon+\delta).
\end{split}
\end{align}
Next observe that for $z_s +\varepsilon\leq z\leq z_n -\varepsilon$, with $\varepsilon$ fixed, formula \eqref{ealp} implies that $\alpha_z =(\bar{u}+\bar{v})_z +O(\rho)$ so that
\begin{align}\label{184y9gfb}
\begin{split}
\alpha(N_{-\varepsilon}^{\delta})-\alpha(S_{\varepsilon}^{\delta})
=&\int_{z_s +\varepsilon}^{z_n -\varepsilon} \alpha_{z}(\delta,z)dz\\
=&\int_{z_s +\varepsilon}^{z_n -\varepsilon} (\bar{u}+\bar{v})_z(\delta,z)dz
+O(\delta)\\
=&(\bar{u}+\bar{v})(N_{-\varepsilon}^{\delta})
-(\bar{u}+\bar{v})(S_{\varepsilon}^{\delta})+O(\delta)\\
=&(\bar{u}+\tilde{v}_n)(N)-(\tilde{u}_s +\bar{v})(S)+O(\varepsilon+\delta).
\end{split}
\end{align}
Combining \eqref{mainform1}, \eqref{mmmmm1}, \eqref{,./}, \eqref{184y9gfb}, and then taking $\delta\rightarrow 0$ followed by $\varepsilon\rightarrow 0$, produces the desired result.

\begin{prop}\label{6.4}
Consider a static bi-axisymmetric solution \eqref{stmetric} of the 5D vacuum Einstein equations. Let $N$ and $S$ be the poles of a spherical negative mass singularity rod that lies between two axis rods, $\Gamma_1$ and $\Gamma_2$, having rod structures $(1,0)$ and $(0,1)$ respectively. Then the difference of logarithmic angle defects is given by
\begin{equation}\label{snms}
b_1 -b_2=\frac{1}{2}\left(\bar{u}(N)-\bar{v}(S)\right)
+\tilde{v}_n(N)-\tilde{u}_s(S).
\end{equation}
\end{prop}

\subsection{Across a ring negative mass singularity}

Finally, we calculate \eqref{mainform2} for a NMS. The arguments are closely related to those of the previous subsection, and so some details will be spared below. Near the poles the following expansions hold
\begin{equation}
u=2\log\rho +\bar{u}, \quad\quad
v=\log\left(r_n +(z-z_n)\right) +\tilde{v}_n \quad \text{ near } N,
\end{equation}
and
\begin{equation}
u=2\log\rho +\bar{u}, \quad\quad
v=\log\left(r_s -(z-z_s)\right) +\tilde{v}_s \quad \text{ near } S,
\end{equation}
for some smooth functions $\bar{u}$, $\tilde{v}_n$, and $\tilde{v}_s$. The expansion of $\alpha$ at the north pole agrees with \eqref{+++++}, while on the other end
\begin{align}\label{-----1}
\begin{split}
\alpha=&\frac{1}{2}\log(1-\cos\theta_s)+\frac{1}{4}\left(2+\frac{\rho^2}{r_s[r_s -(z-z_s)]}\right)\left(\bar{u}-\bar{u}(S)\right)\\
&+\frac{\rho^2}{2r_s[r_s -(z-z_s)]}\left(\tilde{v}_{s}-\tilde{v}_s (S)\right)
+\tilde{c}_s +O(\rho^2)\quad\quad\quad\text{ near }S,
\end{split}
\end{align}
for some constant $\tilde{c}_{s}$. Moreover, equation \eqref{,./} still remains valid while the analogue of \eqref{184y9gfb} becomes
\begin{equation}
\alpha(N_{-\varepsilon}^{\delta})-\alpha(S_{\varepsilon}^{\delta})
=(\bar{u}+\tilde{v}_n)(N)-(\bar{u} +\tilde{v}_s)(S)+O(\varepsilon+\delta).
\end{equation}
Putting these observations together yields the conical singularity propagation formula.

\begin{prop}\label{6.5}
Consider a static bi-axisymmetric solution \eqref{stmetric} of the 5D vacuum Einstein equations. Let $N$ and $S$ be the poles of a ring negative mass singularity rod that lies between two axis rods, $\Gamma_1$ and $\Gamma_2$, both having rod structure $(1,0)$. Then the difference of logarithmic angle defects is given by
\begin{equation}\label{snms}
b_1 -b_2=\frac{1}{2}\left(\bar{u}(N)-\bar{u}(S)\right)
+\tilde{v}_n(N)-\tilde{v}_s(S).
\end{equation}
\end{prop}

\section{Balancing of Signed Masses in 5 Dimensions and the Proof of Theorem \ref{thmsecond}}
\label{sec7} \setcounter{equation}{0}
\setcounter{section}{7}

In this section we will establish Theorem \ref{thmsecond}. The first statement of this result concerns existence for superpositions of positive and negative mass black hole solutions. Indeed, according to the description in terms of Green's functions presented in Section \ref{sec5}, we may define the potential functions $u$ and $v$ as linear combinations of Green's functions for any configuration of signed mass Schwarzschild-Tangherlini and signed
mass static black ring solutions strung along the $z$-axis. With $u$ and $v$, one may solve the quadrature equations \eqref{ealp} for $\alpha$. The resulting spacetime, with metric described in Weyl-Papapetrou coordinates \eqref{metricg1}, is then a bi-axisymmetric solution of the static vacuum Einstein equations. These solutions are regular away from the negative mass singularities, except perhaps for conical singularities on the axes.

\subsection{Balancing two signed spherical masses}

Here we address Theorem \ref{thmsecond} (i), and show that there exists a balanced asymptotically flat configuration of a spherical horizon and a spherical NMS. In particular, an explicit formula for the logarithmic angle defect at each axis will be given in terms of rod data parameters. It will then be shown that for a specific choice of parameters all conical singularities are resolved.

\begin{figure}
	\includegraphics[width=.9\textwidth]{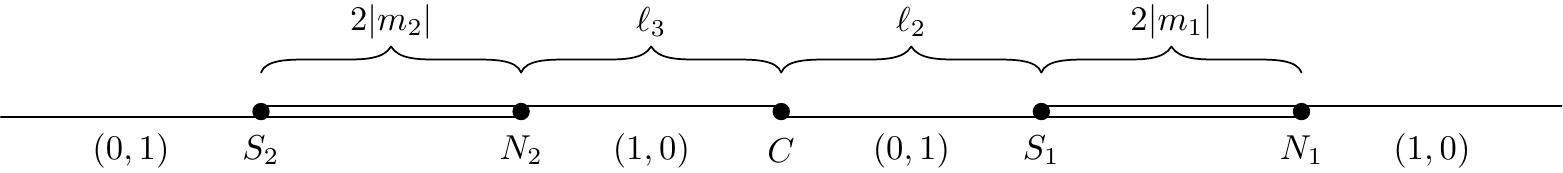}
	\caption{Rod structure for the superposition of two spherical NMSs.}
	\label{fig4}
\end{figure}

We begin by constructing the general rod structure for the two signed masses.
Consider a sequence of four axis rods $\Gamma_i$, $i=1,2,3,4$, with $\Gamma_i$ lying to the north (positive $z$ direction) of $\Gamma_{i+1}$, where $\Gamma_1$, $\Gamma_4$ are semi-infinite rods having rod structures $(1,0)$, $(0,1)$, and $\Gamma_2$, $\Gamma_3$ are finite rods having rod structures $(0,1)$, $(1,0)$. Between $\Gamma_1$ and $\Gamma_2$ is a spherical horizon rod $\Gamma_{h}^1$, and between $\Gamma_3$ and $\Gamma_4$ is a spherical NMS rod $\Gamma_{s}^2$, while the intersection of $\Gamma_2$ and $\Gamma_3$ is a corner point $C=(0,c)$. See Figure \ref{fig4}. The north and south poles of the signed mass rods will be labeled $N_j =(0,n_j)$, $S_j =(0,s_j)$, $j=1,2$ in the $\rho z$-plane, with $s_2 < n_2 < c < s_1 < n_1$. Thus, the total rod structure consists of 6 rods, 4 pole points, and one corner point. The associated potential functions for this rod structure are given by
\begin{equation}
u= G_{\Gamma_1}+G_{\Gamma_3 \cup \Gamma_s^2}
=\log\left(r_{N_1}-(z-n_1)\right)+\log\left[\frac{r_C +(z-c)}{r_{S_2}+(z-s_2)}\right],
\end{equation}
\begin{equation}
v= G_{\Gamma_2}+G_{\Gamma_4 \cup \Gamma_s^2}
=\log\left[\frac{r_{S_1}+(z-s_1)}{r_{C}+(z-c)}\right]
+\log\left(r_{N_2}-(z-n_2)\right),
\end{equation}
where $r_{\cdot}$ denotes Euclidean distance to the relevant point.

We will now compute the propagation of logarithmic angle defects across
the horizon rod $\Gamma_h^1$ and NMS rod $\Gamma_s^2$. Lengths of the two
finite axis rods $\Gamma_2$, $\Gamma_3$ will be denoted by $\ell_2$, $\ell_3$,
and the masses of the horizon/NMS rods will be labeled as $m_1>0$, $m_2<0$. The pole and corner points may now be given the following coordinates on the $z$-axis
\begin{equation}
s_2 =-|m_2|,\quad n_2 =|m_2|,\quad c=\ell_3 +|m_2|,\quad s_1 =\ell_2+\ell_3+|m_2|,\quad
n_1=\ell_2+\ell_3+|m_2|+2m_1.
\end{equation}
Next set
\begin{equation}
\hat{u}_n =u-\log\left(r_{N_1}-(z-n_1)\right)
=\log\left[\frac{r_C +(z-c)}{r_{S_2}+(z-s_2)}\right],
\end{equation}
\begin{equation}
\hat{v}_s =v-\log\left(r_{S_1}+(z-s_1)\right)
=\log\left[\frac{r_{N_2} +(z-n_2)}{r_{C}+(z-c)}\right],
\end{equation}
then according to Proposition \ref{6.2}
\begin{align}\label{462389}
\begin{split}
b_1 -b_2=&\hat{v}_s(S_1)-\hat{u}_n(N_1)+\frac{1}{2}\left(u(S_1)-v(N_1)\right)\\
=&\frac{1}{2}\log\left[\frac{(s_1 -n_2)^2 (n_1 -s_2)^2}
{(s_1-c)(n_1 -c)(s_1 -s_2)(n_1 -n_2)}\right]\\
=&\frac{1}{2}\log\left[\frac{(\ell_2 +\ell_3)^2(\ell_2+\ell_3 +2m_1 +2|m_2|)^2}
{\ell_2(\ell_2+2m_1)(\ell_2+\ell_3+2|m_2|)(\ell_2+\ell_3+2m_1)}\right].
\end{split}
\end{align}
Furthermore let
\begin{equation}
\tilde{u}_s =u-\log\left(r_{S_2}-(z-s_2)\right)
=\log\left[\frac{r_{N_1}-(z-n_1)}{r_C -(z-c)}\right],
\end{equation}
\begin{equation}
\tilde{v}_s=v-\log\left(r_{N_2}+(z-n_2)\right)
=\log\left[\frac{r_C -(z-c)}{r_{S_1}-(z-s_1)}\right],
\end{equation}
\begin{equation}
\bar{u}=u-2\log\rho=-G_{\Gamma_2 \cup \Gamma_h^1}
-G_{\Gamma_4}=-\log\left[\frac{r_C -(z-c)}{r_{N_1}-(z-n_1)}\right]
-\log\left(r_{S_2}+(z-s_2)\right),
\end{equation}
\begin{equation}
\bar{v}=v-2\log\rho=-G_{\Gamma_1 \cup \Gamma_h^1}
-G_{\Gamma_3}=-\log\left(r_{S_1}-(z-s_1)\right)
-\log\left[\frac{r_{N_2}-(z-n_2)}{r_C -(z-c)}\right].
\end{equation}
Then according to Proposition \ref{6.4} we have
\begin{align}\label{=953}
\begin{split}
b_3 -b_4=&\frac{1}{2}\left(\bar{u}(N_2)-\bar{v}(S_2)\right)
+\tilde{v}_n(N_2)-\tilde{u}_s(S_2)\\
=&\frac{1}{2}\log\left[\frac{(n_1 -n_2)(s_1 -s_2)(c-n_2)(c-s_2)}
{(s_1 -n_2)^2 (n_1 -s_2)^2}\right]\\
=&\frac{1}{2}\log\left[\frac{\ell_3(\ell_2+\ell_3+2m_1)(\ell_2+\ell_3+2|m_2|)
(\ell_3+2|m_2|)}{(\ell_2+\ell_3)^2(\ell_2+\ell_3+2m_1+2|m_2|)^2}\right].
\end{split}
\end{align}

In order to show that there is a choice of parameters for which balancing occurs,
we first note that it may be assumed that $b_1=b_4=0$. This is due to the fact that $\alpha$ is defined only up to a constant from \eqref{ealp}, and by appropriately choosing this constant the conical singularity \eqref{angdef} of any given axis rod can be resolved, so without loss of generality we take $b_1=0$. Theorem 2 of \cite{KhuriWeinsteinYamadaP} then shows that due to the asymptotics at spatial infinity we also have $b_4=0$. Thus we only need to show that there are parameters such that the expressions in \eqref{462389} and \eqref{=953} vanish. This may be accomplished by fixing $\ell_2=\ell_3=\ell$, $m_1=|m_2|=m$, and $\mu=m/\ell$ so that
\begin{equation}
-b_2=b_1 -b_2=\frac{1}{2}\log\left[\frac{4(1+2\mu)}{(1+\mu)^2}\right],\quad\quad
b_3=b_3 -b_4=\frac{1}{2}\log\left[\frac{(1+\mu)^2}{4(1+2\mu)}\right].
\end{equation}
In particular, if $\mu=3+2\sqrt{3}$ is the positive root of $(1+\mu)^2=4(1+2\mu)$ then $b_2=b_3=0$. The ADM mass of this balanced solution vanishes by Lemma \ref{admassf1} below.

It should also be mentioned that if the NMS rod $\Gamma_s^2$ is changed to a horizon rod $\Gamma_h^2$ with mass $m_2>0$ so that there are two spherical horizons, or the horizon rod $\Gamma_h^1$ is changed to a NMS rod $\Gamma_s^1$ with mass $m_1<0$ so that there are two spherical NMSs then the propagation of angle defect formula is
\begin{equation}
b_1-b_2=\frac{1}{2}\log\left[\frac{(\ell_2+\ell_3+2|m_1|)^2(\ell_2+\ell_3+2|m_2|)^2}
{\ell_2(\ell_2+\ell_3)(\ell_2+2|m_1|)(\ell_2+\ell_3+2|m_1|+2|m_2|)}\right].
\end{equation}
Since this may be readily shown to always be positive, there can be no balancing with these two configurations. This is to be expected, at least for the case of two horizons, due to the static black hole uniqueness theorem \cite{GibbonsIdaShiromizu1}.

\subsection{Balancing a signed mass black Saturn}

\begin{figure}
\includegraphics[width=.9\textwidth]{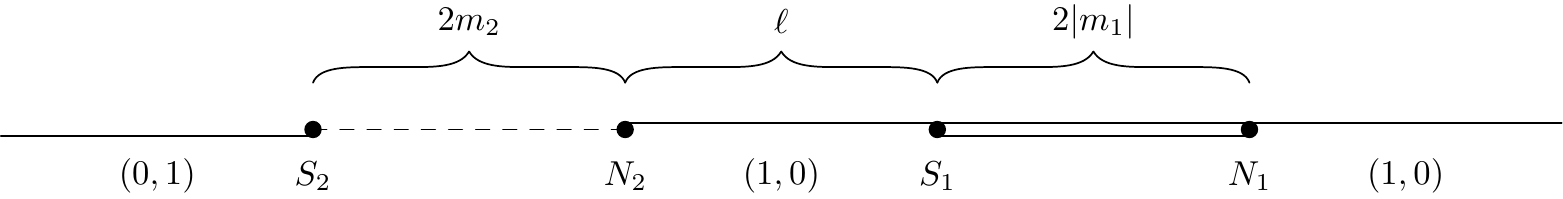}
\caption{Rod structure for the superposition of a ring NMS and spherical horizon.}
\label{fig4.1}
\end{figure}

In the context of Theorem \ref{thmsecond} (i), we balance an asymptotical flat configuration consisting of a ring NMS and a spherical horizon.
Consider a rod structure consisting of three axis rods $\Gamma_i$, $i=1,2,3$, with $\Gamma_i$ lying to the north (positive $z$ direction) of $\Gamma_{i+1}$, where $\Gamma_1$, $\Gamma_3$ are semi-infinite rods having rod structures $(1,0)$, $(0,1)$, and $\Gamma_2$, is a finite rod of length $\ell$ having rod structure $(1,0)$. Between $\Gamma_1$ and $\Gamma_2$ is a ring NMS rod $\Gamma_{s}^1$ of length $2|m_1|$, and between $\Gamma_2$ and $\Gamma_3$ is a spherical horizon rod $\Gamma_{h}^2$ of length $2m_2$; the values $m_1<0$ and $m_2>0$ indicate the masses of these rods. See Figure \ref{fig4.1}. The north and south poles of the signed mass rods will be labeled $N_j =(0,n_j)$, $S_j =(0,s_j)$, $j=1,2$ in the $\rho z$-plane, with $s_2 < n_2 < s_1 < n_1$. The potential functions for this rod structure are
\begin{equation}
u= G_{\Gamma_1\cup \Gamma_s^1 \cup \Gamma_2}
=\log\left(r_{N_2}-(z-n_2)\right),
\end{equation}
\begin{equation}
v= G_{\Gamma_s^1}+G_{\Gamma_3}
=\log\left[\frac{r_{N_1}+(z-n_1)}{r_{S_1}+(z-s_1)}\right]
+\log\left(r_{S_2}+(z-s_2)\right).
\end{equation}

The propagation of logarithmic angle defect across the ring NMS rod $\Gamma_s^1$ may be computed with Proposition \ref{6.5}. In order to
accomplish this label the pole points with the following coordinates on the $z$-axis
\begin{equation}
s_2 =-m_2,\quad n_2 =m_2,\quad s_1=\ell+m_2,\quad n_1=\ell+2|m_1|+m_2,
\end{equation}
and define the regularized potentials
\begin{equation}
\bar{u}=u-2\log\rho=-\log\left(r_{N_2}+(z-n_2)\right),
\end{equation}
\begin{equation}
\tilde{v}_n =v-\log\left(r_{N_1}+(z-n_1)\right)
=\log\left[\frac{r_{S_2} +(z-S_2)}{r_{S_1}+(z-s_1)}\right],
\end{equation}
\begin{equation}
\tilde{v}_s =v-\log\left(r_{S_1}-(z-s_1)\right)
=\log\left[\frac{r_{S_2} +(z-s_2)}{r_{N_1}-(z-n_1)}\right].
\end{equation}
We then have
\begin{align}\label{120-=}
\begin{split}
b_1 -b_2=&\frac{1}{2}\left(\bar{u}(N_1)-\bar{u}(S_1)\right)
+\tilde{v}_n(N_1)-\tilde{v}_s(S_1)\\
=&\frac{1}{2}\log\left[\frac{(s_1-n_2)(n_1-s_2)^2}
{(n_1-n_2)(s_1-s_2)^2}\right]\\
=&\frac{1}{2}\log\left[\frac{\ell(\ell+2|m_1|+2m_2)^2}
{(\ell+2|m_1|)(\ell+2m_2)^2}\right]\\
=&\frac{1}{2}\log\left[\frac{(1+\mu_1 +\mu_2)^2}{(1+\mu_1)(1+\mu_2)^2}\right],
\end{split}
\end{align}
where $\mu_1=2|m_1|/\ell$ and $\mu_2=2m_2/\ell$. As in the previous example we may assume that $b_1=b_3=0$, and so it remains to show that \eqref{120-=} vanishes for a choice of parameters. Indeed, this is the case if $\mu_2=\sqrt{1+\mu_1}$ since then
\begin{equation}
(1+\mu_1 +\mu_2)^2 -(1+\mu_1)(1+\mu_2)^2=\mu_1(1+\mu_1-\mu_2^2)=0.
\end{equation}
Therefore, for any choice of $\ell$ and $m_1<0$, we may set $2m_2=\sqrt{\ell^2 +2|m_1|\ell}$ to achieve $b_2=0$. Furthermore, note that
the ADM mass of this balanced solution is $\tfrac{3\pi}{2}(m_1 +m_2)$ by Lemma \ref{admassf1} below, and thus for appropriate choices of $\ell$ and $m_1$ the ADM mass can be positive, negative, or zero.

\subsection{Balancing two spherical horizons surrounding a spherical NMS}

In this and the next subsection, we will present examples to confirm Theorem \ref{thmsecond} (ii). Namely, here we show that there exists a balanced asymptotically flat configuration of a spherical NMS lying between two spherical horizons. As before, an explicit formula for the logarithmic angle defect at each axis will be given in terms of rod data parameters, which will then be chosen to resolve all conical singularities.

\begin{figure}
	\includegraphics[width=.9\textwidth]{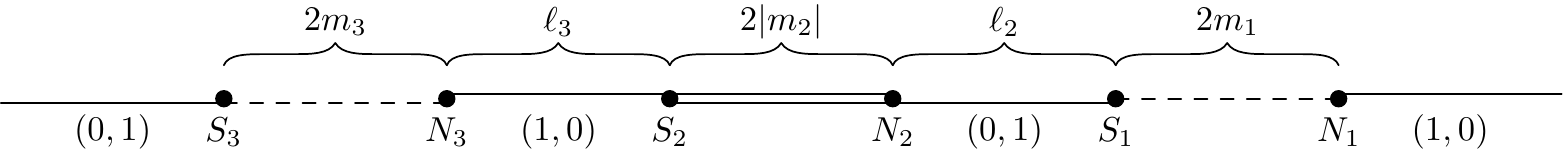}
	\caption{Rod structure for the superposition of two spherical horizons surrounding a spherical NMS.}
	\label{fig5}
\end{figure}

Consider the following rod structure. Let $\Gamma_i$, $i=1,2,3,4$
be a sequence of four axis rods with $\Gamma_i$ lying to the north (positive $z$ direction) of $\Gamma_{i+1}$, where $\Gamma_1$, $\Gamma_4$ are semi-infinite rods having rod structures $(1,0)$, $(0,1)$, and $\Gamma_2$, $\Gamma_3$ are finite rods of lengths $\ell_2$, $\ell_3$ having rod structures $(0,1)$, $(1,0)$. Between $\Gamma_1$ and $\Gamma_2$ is a horizon rod $\Gamma_{h}^1$ of length $2m_1$, between $\Gamma_2$ and $\Gamma_3$ is a NMS rod $\Gamma_s^2$ of length $2|m_2|$, and between $\Gamma_3$ and $\Gamma_4$ is a horizon rod $\Gamma_{h}^3$ of length $2m_3$; as usual $m_i$ represent the signed masses of the horizons/NMSs. See Figure \ref{fig5}. The north and south poles of the signed mass rods are labeled $N_j =(0,n_j)$, $S_j =(0,s_j)$, $j=1,2,3$ in the $\rho z$-plane, with $s_3< n_3< s_2 < n_2 < s_1 < n_1$. Thus, the total rod structure consists of 7 rods and 6 pole points. The associated potential functions for this rod structure are given by
\begin{equation}
u= G_{\Gamma_1}+G_{\Gamma_3 \cup \Gamma_s^2}
=\log\left(r_{N_1}-(z-n_1)\right)+\log\left[\frac{r_{N_2} +(z-n_2)}{r_{N_3}+(z-n_3)}\right],
\end{equation}
\begin{equation}
v= G_{\Gamma_2 \cup \Gamma_s^2}+G_{\Gamma_4}
=\log\left[\frac{r_{S_1}+(z-s_1)}{r_{S_2}+(z-s_2)}\right]
+\log\left(r_{S_3}+(z-s_3)\right).
\end{equation}

We will now compute the propagation of logarithmic angle defects across
the horizon rod $\Gamma_h^1$. Label the $z$-coordinates of the pole points in the following way
\begin{equation}\label{label1}
s_3 =-m_3,\quad n_3 =m_3,\quad s_2=\ell_3 +m_3,\quad n_2=\ell_3+2|m_2|+m_3,
\end{equation}
\begin{equation}\label{label2}
s_1=\ell_2+\ell_3+2|m_2|+m_3,\quad\quad n_1 = \ell_2 +\ell_3 +2m_1 +2|m_2|
+m_3,
\end{equation}
and set
\begin{equation}
\hat{u}_n =u-\log\left(r_{N_1}-(z-n_1)\right)
=\log\left[\frac{r_{N_2} +(z-n_2)}{r_{N_3}+(z-n_3)}\right],
\end{equation}
\begin{equation}
\hat{v}_s =v-\log\left(r_{S_1}+(z-s_1)\right)
=\log\left[\frac{r_{S_3} +(z-s_3)}{r_{S_2}+(z-s_2)}\right].
\end{equation}
Then according to Proposition \ref{6.2}
\begin{align}\label{4623891}
\begin{split}
b_1 -b_2=&\hat{v}_s(S_1)-\hat{u}_n(N_1)+\frac{1}{2}\left(u(S_1)-v(N_1)\right)\\
=&\frac{1}{2}\log\left[\frac{(s_1 -s_3)^2 (n_1 -n_3)^2 (s_1 -n_2)(n_1 -s_2)}
{(s_1-s_2)^2(n_1 -n_2)^2(s_1 -n_3)(n_1 -s_3)}\right]\\
=&\frac{1}{2}\log\left[\frac{\ell_2 (\ell_2 +2m_1 +2|m_2|)(\ell_2+\ell_3 +2m_1 +2|m_2|)^2(\ell_2 +\ell_3 +2|m_2|+2m_3)^2}
{(\ell_2+2m_1)^2 (\ell_2+2|m_2|)^2(\ell_2+\ell_3+2|m_2|)(\ell_2+\ell_3+2m_1 +2|m_2|+2m_3)}\right].
\end{split}
\end{align}

In order to show that there is a choice of parameters for which balancing occurs, we note that as in the previous example it may be assumed that $b_1=b_4=0$. Next observe that by setting $\ell_2=\ell_3=\ell$ and $m_1=m_3$, the rod configuration admits a reflection symmetry across the line $z=\frac{n_2 +s_2}{2}$ passing through the mid point of the NMS rod $\Gamma_s^2$. Consequently the functions $u$, $v$, and $\alpha$ also admit this symmetry and relieving the conical singularity along $\Gamma_2$ implies that it is also relieved along $\Gamma_3$. Thus, it is sufficient to establish
that the expressions in \eqref{4623891} vanishes with an appropriate selection of symmetric parameters. To accomplish this define $\mu_1=m_1 /\ell_2$, $\mu_2=|m_2|/\ell_2$ and note that
\begin{equation}
b_1 -b_2=\frac{1}{2}\log\left[\frac{4(1+\mu_1 +\mu_2)^4(1+2\mu_1 +2\mu_2)}
{(1+2\mu_1)^2(1+2\mu_2)^2(1+\mu_1)(1+2\mu_1+\mu_2)}\right].
\end{equation}
For sufficiently small $\mu_1$ and $\mu_2$ we find $b_1 -b_2 >0$. On the other hand, for sufficiently large $\mu_1 =\mu_2$ we have $b_1 -b_2<0$. Therefore, the intermediate value property yields a choice of $\mu_1 = \mu_2$ such that $b_1 -b_2=0$. In light of Lemma \ref{admassf1} this balanced solution has ADM mass $\tfrac{3\pi}{2}\left(m_1 +m_2 +m_3\right)$, which is positive due to the symmetry assumptions placed on the rod structure.

\subsection{Balancing a ring horizon surrounded by a spherical and ring NMS}

\begin{figure}
	\includegraphics[width=.9\textwidth]{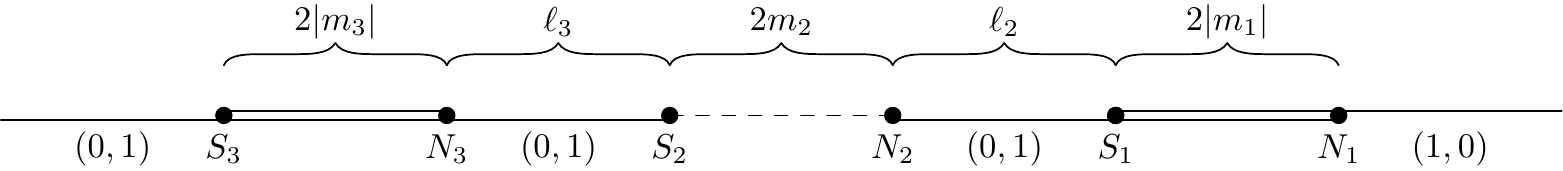}
	\caption{Rod structure for the superposition of a ring horizon
surrounded by a spherical and ring NMS.}
	\label{fig6}
\end{figure}

Consider the following rod structure. As in the previous example there are four axis rods: $\Gamma_1$ having rod structure $(1,0)$, and $\Gamma_i$, $i=2,3,4$ having rod structure $(0,1)$. The first and fourth are semi-infinite, while $\Gamma_2$ and $\Gamma_3$ are finite of length $\ell_2$ and $\ell_3$. Between $\Gamma_1$ and $\Gamma_2$ is a spherical NMS rod $\Gamma_{s}^1$ of length $2|m_1|$, between $\Gamma_2$ and $\Gamma_3$ is a ring horizon rod $\Gamma_h^2$ of length $2m_2$, and between $\Gamma_3$ and $\Gamma_4$ is a ring NMS rod $\Gamma_{s}^3$ of length $2|m_3|$, where $m_i$ represent the signed masses of the horizons/NMSs. See Figure \ref{fig6}. With pole points labeled in the usual fashion, we find that the potential functions associated with this rod structure are given by
\begin{equation}
u= G_{\Gamma_1 \cup\Gamma_s^1}+G_{\Gamma_s^3}
=\log\left(r_{S_1}-(z-s_1)\right)+\log\left[\frac{r_{N_3} +(z-n_3)}{r_{S_3}+(z-s_3)}\right],
\end{equation}
\begin{equation}
v= G_{\Gamma_2 \cup \Gamma_s^1}+G_{\Gamma_3 \cup \Gamma_s^3 \cup\Gamma_4}
=\log\left[\frac{r_{N_1}+(z-n_1)}{r_{N_2}+(z-n_2)}\right]
+\log\left(r_{S_2}+(z-s_2)\right).
\end{equation}

We will first compute the propagation of logarithmic angle defects across
the spherical NMS rod $\Gamma_s^1$. Let the $z$-coordinates of the pole points be labeled as in \eqref{label1}, \eqref{label2} with $m_1$, $|m_2|$, $m_3$ replaced by $|m_1|$, $m_2$, $|m_3|$, and observe that the relevant regularized potentials are
\begin{equation}
\hat{v}_n =v-\log\left(r_{N_1}+(z-n_1)\right)
=\log\left[\frac{r_{S_2} +(z-s_2)}{r_{N_2}+(z-n_2)}\right],
\end{equation}
\begin{equation}
\hat{u}_s =u-\log\left(r_{S_1}-(z-s_1)\right)
=\log\left[\frac{r_{N_3} +(z-n_3)}{r_{S_3}+(z-s_3)}\right],
\end{equation}
\begin{equation}
\bar{u}=u-2\log\rho=-G_{\Gamma_2 \cup\Gamma_h^2 \cup\Gamma_3}
-G_{\Gamma_4}=-\log\left[\frac{r_{S_1}+(z-s_1)}{r_{N_3}+(z-n_3)}\right]
-\log(r_{S_3}+(z-s_3)),
\end{equation}
\begin{equation}
\bar{v}=v-2\log\rho=-G_{\Gamma_1}-G_{\Gamma_h^2}
=-\log(r_{N_1}-(z-n_1))-\log\left[\frac{r_{N_2}+(z-n_2)}{r_{S_2}+(z-s_2)}\right].
\end{equation}
Then according to Proposition \ref{6.4}
\begin{align}\label{462389100}
\begin{split}
b_1 -b_2=&\frac{1}{2}\left(\bar{u}(N_1)-\bar{v}(S_1)\right)
+\tilde{v}_n(N_1)-\tilde{u}_s(S_1)\\
=&\frac{1}{2}\log\left[\frac{(n_1 -n_3) (s_1 -n_2)(n_1 -s_2)^2(s_1 -s_3)^2}
{(n_1-s_3)(s_1 -s_2)(n_1 -n_2)^2(s_1 -n_3)^2}\right]\\
=&\frac{1}{2}\log\left[\frac{\ell_2 (\ell_2+\ell_3 +2|m_1| +2m_2)(\ell_2 +2|m_1| +2m_2)^2(\ell_2 +\ell_3 +2m_2+2|m_3|)^2}
{(\ell_2+\ell_3+2|m_1| +2m_2+2|m_3|)(\ell_2+2m_2)(\ell_2+2|m_1|)^2 (\ell_2+\ell_3+2m_2)^2}\right].
\end{split}
\end{align}

Next consider the propagation of logarithmic angle defect across the ring NMS rod $\Gamma_s^3$. The relevant regularized potential functions are given by
\begin{equation}
\hat{n}_n =u-\log\left(r_{N_3}+(z-n_3)\right)
=\log\left[\frac{r_{S_1} -(z-s_1)}{r_{S_3}+(z-s_3)}\right],
\end{equation}
\begin{equation}
\hat{u}_s =u-\log\left(r_{S_3}-(z-s_3)\right)
=\log\left[\frac{r_{S_1} -(z-s_1)}{r_{N_3}-(z-n_3)}\right],
\end{equation}
\begin{equation}
\bar{v}=v-2\log\rho=-G_{\Gamma_s^2}
-G_{\Gamma_1}=-\log\left[\frac{r_{S_2}-(z-s_2)}{r_{N_2}-(z-n_2)}\right]
-\log(r_{N_1}-(z-n_1)).
\end{equation}
It then follows from Proposition \ref{6.5} that
\begin{align}\label{4623891001}
\begin{split}
b_3 -b_4=&\frac{1}{2}\left(\bar{v}(N_3)-\bar{v}(S_3)\right)
+\tilde{u}_n(N_3)-\tilde{u}_s(S_3)\\
=&\frac{1}{2}\log\left[\frac{(n_2 -n_3) (s_2 -s_3)(n_1 -s_3)(s_1 -n_3)^2}
{(s_2-n_3)(n_1 -n_3)(n_2 -s_3)(s_1 -s_3)^2}\right]\\
=&\frac{1}{2}\log\left[\frac{(\ell_3+2m_2)(\ell_3+2|m_3|) (\ell_2+\ell_3 +2|m_1| +2m_2+2|m_3|)(\ell_2 +\ell_3 +2m_2)^2}
{\ell_3(\ell_2+\ell_3+2|m_1| +2m_2)(\ell_3+2m_2+2|m_3|)(\ell_2+\ell_3+2m_2+2|m_3|)^2}\right].
\end{split}
\end{align}

As in the previous example we may assume that $b_1=b_4=0$. In order to show that there is a choice of parameters for which $b_2=b_3=0$, let $\ell_2=\ell_3=\ell$ and set $\mu_1=|m_1|/\ell$, $\mu_2=m_2/\ell$, $\mu_3=|m_3|/\ell$. Then the conical singularity propagation formulas may be rewritten as
\begin{equation}
-b_2=b_1-b_2=-\frac{1}{2}\log\left[\frac{(1+\mu_1+\mu_2+\mu_3)(1+\mu_2)^2}
{(1+\mu_1+\mu_2)(1+\mu_2+\mu_3)^2}\right]
+\frac{1}{2}\log\left[\frac{(1+2\mu_1+2\mu_2)^2}{(1+2\mu_2)(1+2\mu_1)^2}\right],
\end{equation}
\begin{equation}
b_3=b_3-b_4=\frac{1}{2}\log\left[\frac{(1+\mu_1+\mu_2+\mu_3)(1+\mu_2)^2}
{(1+\mu_1+\mu_2)(1+\mu_2+\mu_3)^2}\right]
+\frac{1}{2}\log\left[\frac{(1+2\mu_2)(1+2\mu_3)}{(1+2\mu_2+2\mu_3)}\right].
\end{equation}
Then $b_2=b_3=0$ is equivalent to
\begin{equation}\label{3by3}
\underbrace{\frac{(1+2\mu_1+2\mu_2)^2}{(1+2\mu_2)(1+2\mu_1)^2}}_{I}
=\underbrace{\frac{(1+\mu_1+\mu_2+\mu_3)(1+\mu_2)^2}
{(1+\mu_1+\mu_2)(1+\mu_2+\mu_3)^2}}_{II}
=\underbrace{\frac{(1+2\mu_2+2\mu_3)}{(1+2\mu_2)(1+2\mu_3)}}_{III}.
\end{equation}
Notice that $I=III$ is satisfied if
\begin{equation}\label{mu3}
\mu_3=\frac{(1+2\mu_1)^2-4(1+2\mu_1)-4\mu_2}{4(1+2\mu_1)+4\mu_2}>0.
\end{equation}
In order to also achieve $I=II$, consider a curve in parameter space $\tau\rightarrow (\mu_1(\tau),\mu_2(\tau),\mu_3(\tau))$ for  $\tau\in[0,\varepsilon]$, where $\varepsilon>0$ is small.
The component functions are as follows: $\mu_2(\tau)=\tau$, $\mu_3(\tau)$ is defined by \eqref{mu3}, and $\mu_1(\tau)$ is chosen to obtain
\begin{equation}
(1+2\mu_1(\tau))^2-4(1+2\mu_1(\tau))-4\mu_2(\tau)=4\varepsilon^2\quad\quad
\Rightarrow\quad\quad \mu_1(\tau)=\frac{1}{2}+\sqrt{1+\tau+\varepsilon^2}.
\end{equation}
This prescription for the curve implies
\begin{equation}
\mu_3(\tau)=\frac{1}{4}\varepsilon^2+O(\tau\varepsilon^2+\varepsilon^4)>0,\quad
\log I(\tau)=-\tau+O(\tau\varepsilon^2+\tau^2),\quad
\log II(\tau)=-\frac{2}{5}\varepsilon^2+O(\tau\varepsilon^2+\varepsilon^4).
\end{equation}
Thus, along the curve we have that $I(\tau)=III(\tau)$ for all small $\tau$, and
\begin{equation}
\log II(0)=-\frac{2}{5}\varepsilon^2+O(\varepsilon^3) <\log I(0)=0,
\end{equation}
\begin{equation}
\log II(\varepsilon^{3/2})=-\frac{2}{5}\varepsilon^2+O(\varepsilon^3) >
\log I(\varepsilon^{3/2})=-\varepsilon^{3/2}+O(\varepsilon^3),
\end{equation}
if $\varepsilon$ is appropriately small. It follows that there is a time $\tau\in(0,\varepsilon^{3/2})$ such that $I(\tau)=II(\tau)$. This yields a solution of \eqref{3by3}, and consequently the balancing of all axis rods.
Moreover, according to Lemma \ref{admassf1} the ADM mass of the balanced solution is $\tfrac{3\pi}{2}\left(m_1 +m_2 +m_3\right)$, which is negative since both $m_2$ and $m_3$ are small compared to $m_1 <0$.

\subsection{The ADM mass of superpositioned signed masses in 5 dimensions}
\label{massec5}

Here the ADM mass of asymptotically flat, bi-axisymmetric, static vacuum spacetimes
having both horizons and negative mass singularities is computed in terms of the
horizon/NMS rod lengths. Recall that the total mass is given by the Komar integral
\begin{equation}\label{ADM-komar-5d}
m= -\frac{3}{32\pi} \int_{S_\infty^3} \star d\xi,
\end{equation}
where $\xi$ denotes the dual 1-form to the timelike Killing field $\partial_t$, and
$S_\infty^{3}$ denotes a limit of 3-dimensional coordinate spheres $S_r^3$ in the asymptotically flat end as $r\rightarrow\infty$. Write $U=u+v -2\log\rho$ so that the spacetime metric \eqref{metricg1} becomes
\begin{equation}
g = - e^{-U} dt^2 + e^u  d \phi^2 +e^v d\psi^2 +  e^{2 \alpha} (d \rho^2 + dz^2).
\end{equation}
We find that
\begin{equation}\label{polarform1}
\star d \xi = \rho  e^{2 \alpha - U} (- U_\rho dz \wedge d\phi\wedge d\psi + U_z d \rho \wedge d \phi \wedge d\psi)=U_r e^{2\alpha -U} r^3 \sin(2\theta) d\theta \wedge d\phi \wedge d\psi,
\end{equation}
where the polar coordinates are defined by $\rho=r^2\sin2\theta$, $z=r^2\cos2\theta$ for $\theta\in[0,\pi/2]$. Since $U$ is harmonic, an expansion in spherical harmonics in the asymptotically flat end yields $U=\frac{c}{r^2}+O(r^{-3})$, for some constant $c$. Moreover, as in the 4-dimensional case treated in Section \ref{massec1}, the quadrature equations for $\alpha$ imply that $|\nabla(2\alpha-U)|=O(r^{-1})$. It follows that
\begin{equation}
\star d\xi=\left(-\frac{2c}{r^3}+O(r^{-4})\right)\left(1+O(r^{-1})\right)r^3 \sin(2\theta) d\theta\wedge d\phi\wedge d\psi.
\end{equation}
We may now evaluate the Komar mass integral to find $m=\tfrac{3\pi c}{4}$, or rather
\begin{equation}\label{massubar1}
U=\frac{4m}{3\pi r^2}+O(r^{-3}).
\end{equation}

Consider now a general asymptotically flat rod structure with horizon rods $\Gamma_{h}^i$, $i=1,\ldots, i_0$ having masses $m_i>0$, and NMS rods $\Gamma_s^j$, $j=1,\ldots,j_0$ having masses $m_j <0$, then
\begin{equation}\label{baruu1}
U=-\sum_{i=1}^{i_0} G_{\Gamma_h^i} +\sum_{j=1}^{j_0} G_{\Gamma_s^j}.
\end{equation}
If the $z$-components of the north and south poles of the horizon/NMS rods are labelled $n_i$, $s_i$, $n_j$, $s_j$, then the Green's functions have the expansions at infinity
\begin{equation}
G_{\Gamma_h^i}=\frac{s_i - n_i}{r^2}+O(r^{-3}),\quad\quad\quad
G_{\Gamma_s^j}=\frac{s_j -n_j}{r^2}+O(r^{-3}).
\end{equation}
We have thus obtained the following formula for the ADM mass.

\begin{lemma}\label{admassf1}
Consider a 5-dimensional, asymptotically flat, bi-axisymmetric, static vacuum configuration consisting of the superposition of signed mass Schwarzschild-Tangherlini and signed mass black ring solutions having horizon masses $\tfrac{3\pi}{2}m_i>0$, $i=1,\ldots, i_0$, and NMS masses $\tfrac{3\pi}{2}m_j<0$, $j=1,\ldots, j_0$. If the corresponding horizon and NMS rods have $z$-components in Weyl-Papapetrou coordinates given by $n_i$, $s_i$, $n_j$, $s_j$ respectively, then the ADM mass of this configuration is
\begin{equation}
m=\frac{3\pi}{4}\left(\sum_{i=1}^{i_0}(n_i -s_i)-\sum_{j=1}^{j_0}(n_j -s_j)\right)
=\frac{3\pi}{2}\left(\sum_{i=1}^{i_0}m_i +\sum_{j=1}^{j_0}m_j\right).
\end{equation}
\end{lemma}

\end{document}